\documentclass[12pt, aip,cha,graphicx,amsmath,amssymb,reprint,onecolumn]{revtex4-1}
\usepackage{amssymb, graphics, epsfig, color, ulem, float,graphicx, suppenv}
\usepackage{amsmath}
\begin{document}
\titlepage
\title{Time irreversibility, entropy production and effective temperature are independently regulated in the actin cortex of living cells}
\author{N Narinder}
\affiliation{Cluster of Excellence Physics of Life, Technische Universit\"at Dresden, Dresden, Germany}
\affiliation{School of Science, Technische Universit\"at Dresden, Dresden, Germany}
\author{Elisabeth Fischer-Friedrich\thanks{corresponding author}}
\email{Corresponding author: elisabeth.fischer-friedrich@tu-dresden.de}
\affiliation{Cluster of Excellence Physics of Life, Technische Universit\"at Dresden, Dresden, Germany}
\affiliation{School of Science, Technische Universit\"at Dresden, Dresden, Germany}
\date{November 2024}

\begin{abstract} 
Living cells exhibit non-equilibrium dynamics emergent from the intricate interplay between molecular motor activity and its viscoelastic cytoskeletal matrix.  The deviation from thermal equilibrium can be quantified through frequency-dependent effective temperature or time-reversal symmetry breaking quantified e.g. through the Kullback-Leibler divergence. 
Here, we investigate the fluctuations of an AFM tip embedded within the active cortex of mitotic human cells with and without perturbations that reduce cortex activity through inhibition of material turnover or motor proteins. While inhibition of motor activity significantly reduces both effective temperature and time irreversibility, inhibited material turnover leaves the effective temperature largely unchanged but lowers the time irreversibility and entropy production rate. Our experimental findings in combination with a minimal model highlight that time irreversibility, effective temperature and entropy production rate can follow opposite trends in active living systems, challenging in particular the validity of effective temperature as a proxy for the distance from thermal equilibrium.
Furthermore, we propose that the strength of thermal noise and the occurrence of time-asymmetric deflection spikes in the dynamics of regulated observables are inherently coupled in living systems, revealing a previously unrecognized link between entropy production and time irreversibility.
\end{abstract}

\maketitle
Mechanics and statistical physics describe the laws governing the motion of objects at mesoscopic and macroscopic scales in thermodynamic equilibrium. Living cells are paradigmatic non-equilibrium systems due to their continuous consumption and dissipation of energy. To characterize the dynamics of living cells, methods that quantify deviations from equilibrium are essential. In equilibrium, the fluctuation-dissipation theorem (FDT) relates the response function of a physical variable to the strength of its fluctuations. In active systems, deviations from this relation provide a direct verification of non-equilibrium activity. A prevalent approach to quantify this violation is the definition of an effective temperature $T_{\mathrm{eff}} $ whose excess over the ambient temperature reflects the degree to which fluctuations surpass thermal levels.  
Additionally, the asymmetry of active dynamics under time reversal, e.g.  quantified by the Kullback-Leibler divergence (KLD), serves as a second important indicator of activity. 

Previous biophysics research has used effective temperature $T_{\mathrm{eff}}$  or time irreversibility as a measure of deviation from equilibrium implying a monotonic increase of these measures with entropy production rate, see e.g. \cite{ben-isaac_effective_2011, bohec_distribution_2019, mizuno2007nonequilibrium, schlosser_force_2015, hurst_intracellular_2021,martinez_inferring_2019, roldan2021quantifying, martin_comparison_2001, muenker_accessing_2024}. However, direct experimental validation of this assumption remains elusive. To fill this gap, we provide here a comparative study of both, effective temperature and time irreversibility, in relation to entropy production rate applied in an exemplary manner to the active dynamics of the actin cortex - a thin biopolymer network underneath the plasma membrane of animal cells. 

\section{Active fluctuations of cortex-embedded AFM tips}
\begin{figure}[!]
    \centering
    \includegraphics[scale=0.60]{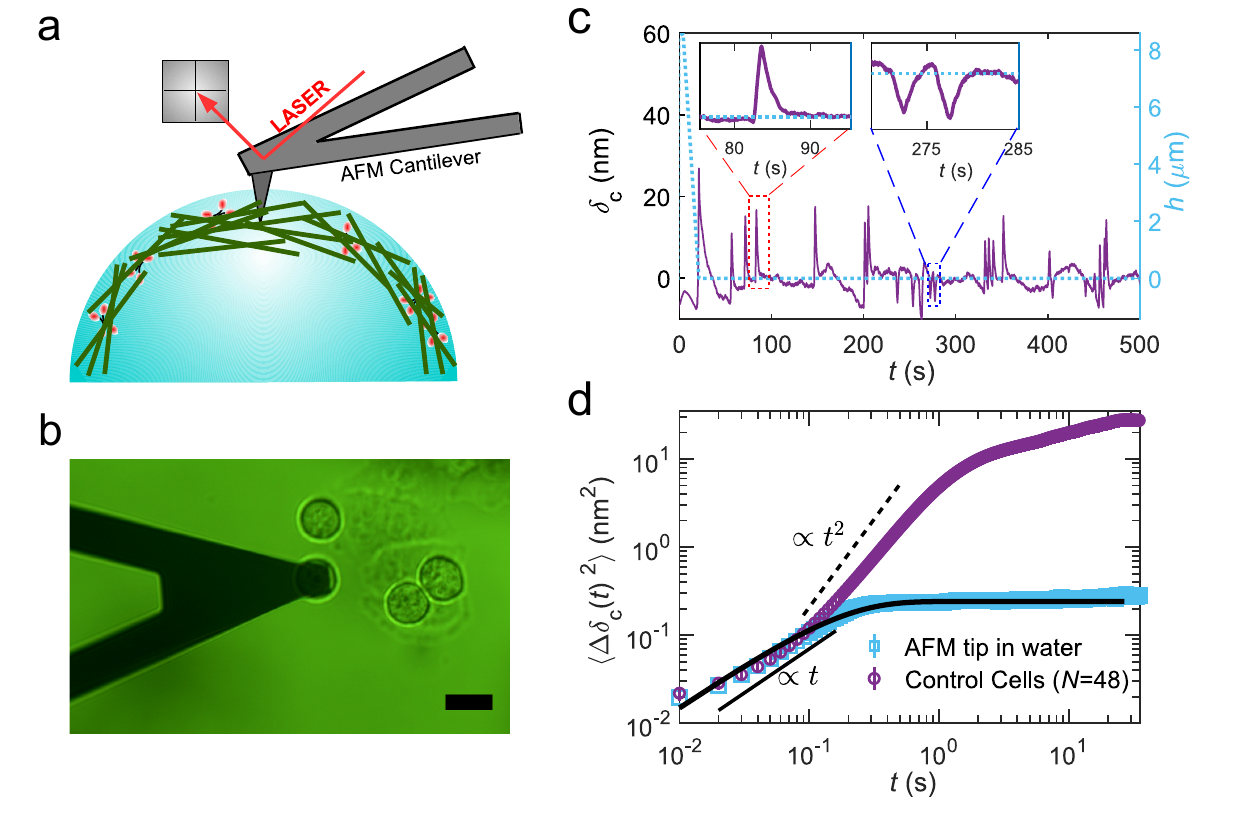}
    \caption{Probing active cortex fluctuations with AFM. (a) Schematic illustration of the experiment (green: actin polymers, red: motor proteins). 
    (b) Microscopy image showing cells and an AFM cantilever during the measurement (bottom view). The pyramidal AFM cantilever tip is embedded in the cortex of a HeLa cell in mitotic arrest from the top. Scale bar: $20~\mathrm{\mu m}$. (c) Left: A typical experimental time progression of the tip position $\delta_\mathrm{c}(t)$ embedded within the cortex. 
    Right: Corresponding piezo-controlled hinge height $h(t)$ as a function of time.
    For simplicity, final hinge height and the  mean of $\delta_\mathrm{c}(t)$ were set to zero. 
    Left Inset: Enlarged view of an active force burst directed towards the cantilever displaying a sudden increase of the tip position followed by an exponential-like relaxation. 
    Right Inset: Expanded view around an active force burst directed towards the interior of the cell with an approximately time-symmetric shape. 
    (d) Time dependence of the measured mean-squared displacement of a tip inserted within the cortex of mitotic cells (purple circles) and for the tip suspended in aqueous medium (blue square). The black solid curve represents a fit to Eqn.~\eqref{eq:langevin_free_tip} with $\kappa=0.0355~\mathrm{N/m}$ and $\gamma=0.0056~\mathrm{Ns/m}$.}
    \label{fig:panel_1}
\end{figure}

To experimentally assess the active mechanics of the cell
cortex, we chose to work with HeLa cells arrested in mitosis. Such cells are particularly suitable for our study  as they naturally adopt an almost spherical shape with a strongly activated molecular motor activity in the actin cortex \cite{fischer2014quantification, chugh2017}. Further, we arrested the cells in mitosis to prevent temporal changes of material properties due to progression of cells through the cell cycle, see Materials and Methods.  During our measurements, we embedded the tip of a pyramidal AFM cantilever in the cortex.
A schematic sketch representing the experimental assay is depicted in Fig.~\ref{fig:panel_1}\,(a). 
A snapshot (bottom view) of a mitotic cell being probed by the AFM cantilever is shown in Fig.~\ref{fig:panel_1}\,(b). Since the embedded AFM tip gets mechanically coupled to the cortex, its deflection serves as a readout of the local active fluctuations within the cortical layer. 

In Fig.~\ref{fig:panel_1}\,(c), we present a representative time evolution of the cantilever hinge height, $h$, alongside the vertical tip deflection, $\delta_\mathrm{c} = F / \kappa$, where $F$ denotes the measured AFM force and $\kappa$ is the cantilever's spring constant. 
Once the set-point force is reached ($t \approx 20$~s), the hinge height remains constant, while the tip deflection exhibits an exponential-like relaxation — a characteristic response of a viscoelastic material~\cite{khalilgharibi2019stress}. 
Subsequently, the tip position fluctuates around $\delta_\mathrm{c} = 0$, driven by a combination of thermal and athermal forces acting on the system.

Strikingly, the tip position exhibits pronounced active bursts directed outward from the cell (upward tip motion), with amplitudes of approximately $A_{u} \approx 10~\mathrm{nm}$ and a subsequent relaxation which is captured by an exponential decay with timescale of $\tau \approx 1.5~\mathrm{s}$, see Fig.~\ref{fig:supp_panel_1_new}\,(a)-(d).
A magnified view of an upward active burst is shown in the left inset of Fig.~\ref{fig:panel_1}\,(c). We note that these bursts are inherently asymmetric in time, see also Fig.~\ref{fig:supp_panel_1_new}\,(c)-(g). 
Moreover, the tip also undergoes substantial deflection events directed toward the cell interior, as shown in the inset of Fig.~\ref{fig:panel_1}\,(c). However, these events have smaller amplitudes $A_d \approx 5~\mathrm{nm}$ and exhibit a less asymmetric shape, see Fig.~\ref{fig:supp_panel_1_new}\,(f) and Fig.~\ref{fig:supp_panel_1_new}\,(l)–(o), respectively. In addition to this, the active nature of the tip fluctuations is also evident from the probability distribution function of tip displacements $P(\Delta\delta_\mathrm{c})$ at various time lags which exhibit strongly non-Gaussian tails~\cite{ben-isaac_modeling_2015}, as shown in Fig.~\ref{fig:supp_panel_2_new}. 

To quantify the active fluctuations described above, we computed the mean squared displacement (MSD) of the tip position, $\langle \Delta \delta_\mathrm{c} (t)^2\rangle=\langle[\delta_\mathrm{c}(t+t_0)-\delta_\mathrm{c}(t_0)]^2\rangle$. 
For AFM cantilevers freely fluctuating in cell culture medium, tip fluctuations can be described as that of an overdamped harmonic oscillator in a viscous medium
\begin{equation}
\langle \Delta \delta_\mathrm{c}^2 (t) \rangle = \frac{2k_\mathrm{B}T}{\kappa} \left( 1 - \exp{\left( -\frac{\kappa t}{\gamma} \right)} \right) \quad,
\label{eq:langevin_free_tip}
\end{equation}
 where $\gamma$ represents the viscous friction coefficient, see Fig.~\ref{fig:panel_1}\,(d), blue curve.
Correspondingly, the MSD converges to a plateau  $2k_\mathrm{B}T/\kappa$ with $\kappa$ approximately the nominal spring constant of the cantilever and $T$ the ambient bath temperature.
Notably, fluctuations of AFM cantilevers embedded in the active cell cortex were strongly enhanced as compared to cantilevers freely immersed in medium, see purple curve in Fig.~\ref{fig:panel_1}\,(d) and Fig.~\ref{fig:supp_panel_3_new}.
Further, the MSD of cortex-immersed tips displays a transition from a diffusive behavior at short times $t\lesssim0.1~\mathrm{s}$ to super-diffusive dynamics for $0.1~\mathrm{s} \lesssim t \lesssim 1~\mathrm{s}$, a typical signature of the active nature of the cortex \cite{fakhri2014high}.
Due to the caged motion of the AFM tip, the MSD converges to a plateau $\langle \Delta \delta_\mathrm{c} (t\rightarrow \infty)^2 \rangle=2k_\mathrm{B}T_\mathrm{eff}/\kappa$, where $T_\mathrm{eff}$ represents the effective temperature in the long-timescale limit\cite{bruno_transition_2009, ghosh_dynamics_2014}.
Using the measured plateau value, we can estimate the effective temperature, $T_\mathrm{eff}$ to be about two orders of magnitude higher than the ambient bath temperature $T\approx 310~\mathrm{K}$ of the cells.

\section{Effective temperature and entropy production}
\begin{figure}[!]
    \centering
    \includegraphics[scale=0.75]{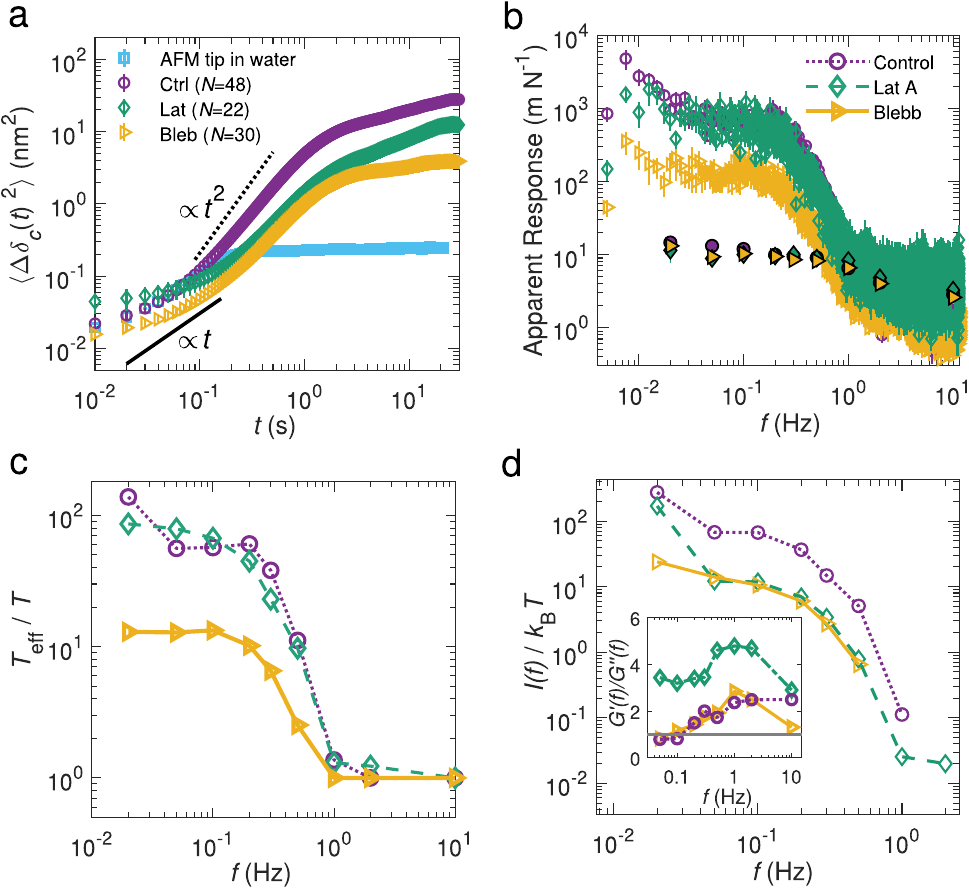}
    \caption{Enhanced effective temperature highlights cortex activity.  
    (a) Experimentally measured MSD $\langle \Delta \delta_\mathrm{c} (t)^2\rangle$ for the tip suspended in aqueous medium at $T=310~\mathrm{K}$ and inserted within the cortex of the mitotic cells under control (purple circles) and pharamacologically treated conditions: $10~\mathrm{\mu M}$ Blebbistatin (yellow triangles) and $25~\mathrm{n M}$ Latruculin-A (green diamonds). (b) Deviation from the fluctuation-dissipation theorem (Eqn.~\eqref{eq:fdt_main}). The open symbols represent $\pi f C(f)/k_\mathrm{B}T$, while the filled symbols correspond to the imaginary part of the response function, $\chi''(f)$. Symbols and color codes are as in panel (a).  
    (c) Effective temperature $T_\mathrm{eff}$ normalized by the ambient temperature as a function of frequency obtained from the violation of the FDT from panel (b) for control (purple circles), treated with  Latrunculin-A (green diamonds) or Blebbistatin (yellow triangles).
    (d) Energy dissipation spectrum $I(f)$ obtained from the fluctuations of AFM cantilever tips embedded in the mitotic actin cortex. Symbols and color codes are as in panel (a). Inset: Experimentally measured ratio of storage modulus to loss modulus, $G'(f)/G''(f)$, as a function of frequency, characterizing the solid-like nature of the actin cortex. The gray curve represents the reference line where $G'(f)/G''(f) = 1$, indicating the transition between predominantly elastic and viscous behavior.}
    \label{fig:panel_2}
\end{figure}

The actin cortex is subject to constant ATP-driven polymer turnover and active forces exerted by associated motor proteins~\cite{salbreux_actin_2012}.
To dissect the effect of these processes on active fluctuations, we applied low doses of Latrunculin-A 
($25~\mathrm{nM}$) to slow down actin polymerization and (-)-Blebbistatin ($10~\mathrm{\mu M}$) to reduce active myosin forces \cite{limouze2004specificity,fischer2016rheology}.
In Fig.~\ref{fig:panel_2}\,(a), we plotted the measured MSD of tip fluctuations subject to the above-mentioned drug treatments along with measurement results of untreated cells (control) and of the bare tip in aqueous medium. 
In comparison to control, the MSD for both treatments exhibits quantitatively distinct long-time behavior with a lowered plateau in the long time limit reflecting reduced cortical activity. 

To better quantify the out-of-equilibrium nature of tip fluctuations in different situations, we investigated the dynamics using the fluctuation-dissipation theorem (FDT) 
\cite{turlier2016equilibrium, landau_statistical_2011} 
\begin{equation}
\frac{C(f)}{2k_\mathrm{B}T}=\frac{\chi''(f)}{2\pi f} \quad,
\label{eq:fdt_main}
\end{equation}
where $f$ is the frequency of fluctuations.
As described in the Materials and Methods section, the FDT relates the power spectral density $C(f)$ of tip fluctuations $\delta_\mathrm{c}(t)$ to the imaginary part of the response function $\chi''(f)$. The latter was determined via active sinusoidal cantilever oscillations at frequencies $f=0.02-10$~Hz, see Materials and Methods. 
The obtained results for $\pi f C(f)/(k_\mathrm{B}T)$ (open symbols) and $\chi''(f)$ (filled symbols)  are plotted in Fig.~\ref{fig:panel_2}\,(b). The violation of the FDT for a specific frequency is marked by the ratio between the two quantities, \textit{i.e.}, $T_\mathrm{eff}(f)/T$. Our experiments clearly show a violation of the FDT for frequencies $f\lesssim 1~\mathrm{Hz}$ as indicated by $T_\mathrm{eff}(f)/T>1$ in this regime, see Fig.~\ref{fig:panel_2}\,(c). This observation confirms the active nature of the measured tip fluctuations within the cortex, suggesting that cell cortex activity predominantly arises from relatively slow biological processes, such as myosin activity, actin polymerization and blebbing kinetics~\cite{charras2006reassembly,fritzsche2013analysis,ruffine2023twofold, mizuno2007nonequilibrium, kovacs_functional_2003, ben-isaac_effective_2011}. For $f\gtrsim 1~\mathrm{Hz}$, the fluctuations are dominated by passive thermal noise. Consistent with the observed MSD, the degree of violation is more pronounced for control cells (purple circles) and cells with turnover inhibition (green diamonds), as compared to cells with inhibited motor activity (yellow triangles). 
Interestingly, the measurement curves of effective temperature versus frequency are well captured by an exponential decay of effective temperature with characteristic timescale of $2/3$~s, see Fig.~\ref{fig:supp_panel_10_new}. 

Earlier research demonstrated that the effective temperature in a viscoelastic fluid is linked to the energy dissipation spectrum $I(f)$ by the Harada-Sasa equality \cite{harada_sasa_2005,deutsch_energy_2006,fodor2016nonequilibrium}
\begin{eqnarray}
I(f)&=&\frac{2k_\mathrm{B}(T_\mathrm{eff}(f)-T)}{1+(G'(f)/G''(f))^2}, \quad.
\label{eq:energy_spectrum}
\end{eqnarray}
The corresponding entropy production rate is $\sigma = \int df I(f)/T$. 
Here, \( G'(f) \) and \( G''(f) \) represent the real and imaginary components of the frequency-dependent complex shear modulus \( G^*(f) \) of the actin cortex, where the real part characterizes the solid-like response and the imaginary part captures the fluid-like response. 
Accordingly, $I(f)$ vanishes for $T_\mathrm{eff}(f)=T$ which corresponds to the thermal equilibrium scenario. Additionally, $I(f)$ depends on the mechanical properties of the surrounding environment characterized by the inverse loss tangent $G'(f)/G''(f)$, which is a measure of solidity. This functional dependence acknowledges that only the fluid-like response of a material gives rise to energy dissipation while  in a purely solid-like material fluctuations merely reflect time-dynamic exchange between kinetic and elastic energy. 

Performing active rheological measurements, see Fig.~\ref{fig:supp_panel_4_new} and Materials and Methods, we determined the complex elastic modulus $G^*(f)$, see Fig.~\ref{fig:supp_panel_5_new} and Materials and Methods. 
Thus, we can calculate $G'(f)/G''(f)$, see inset Fig.~\ref{fig:panel_2}\,(d). Notably, we find  a marked increase of $G'(f)/G''(f)$ for Latrunculin-treated cells where actin material turnover is inhibited. 
Using results for $G'(f)/G''(f)$, we determine $I(f)$ for both control and treated cells, see Fig.~\ref{fig:panel_2}\,(d). 
Due to the more pronounced solid-like nature of Latrunculin-treated cells, dissipation is reduced as compared to Blebbistatin-treated cells.
Thus, while the Latrunculin-treated cells exhibit a larger effective temperature $T_\mathrm{eff}$ compared to the Blebbistatin-treated cells, their dissipation spectra $I(f)$ are similar. 
By integrating the normalized dissipation spectrum over the entire frequency range, we obtain the entropy production rates:  
\(\sigma = 19.58~{k_\mathrm{B}/\mathrm{s}}\) for control cells,  
\(5.51~{k_\mathrm{B}/\mathrm{s}}\) for Latrunculin-treated cells, and  
\(2.94~{k_\mathrm{B}/\mathrm{s}}\) for Blebbistatin-treated cells.
We conclude that upon turnover inhibition (Latrunculin-A treatment), entropy production rate decreases significantly, while the effective temperature remains largely unchanged.

\section{Time irreversibility follows entropy production trends}
\begin{figure}[!]
    \centering
    \includegraphics[scale=0.63]{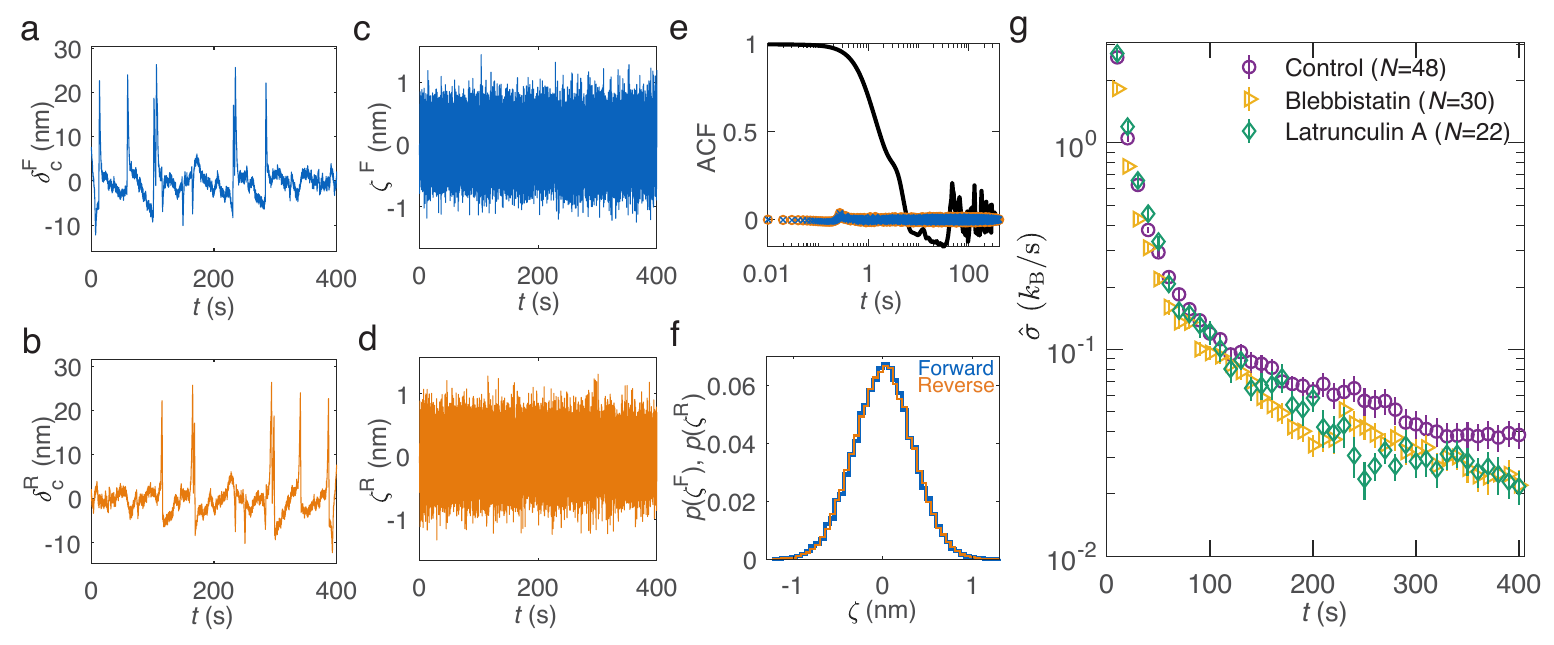}
    \caption{Inhibition of actin turnover and motor activity reduces time irreversibility. (a)-(b) Representative steady-state fluctuations of the AFM tip immersed within the cortical layer of a mitotic HeLa cell, shown for both time forward $\delta_\mathrm{c}^\mathrm{F}$ (a) and time reverse $\delta_\mathrm{c}^\mathrm{R}$ (b) directions. 
    (c)-(d) Corresponding residual time series $\zeta^{\rm F}$ and $\zeta^{\rm R}$ obtained from either signal by subtracting  the autoregression model obtained by fitting the time reverse tip fluctuations. 
    (e) The autocorrelation function for the measured time-forward AFM tip fluctuations (solid black curve), along with the autocorrelation function of the whitened residual signals $\zeta^{\rm F}$ ($\times$) and $\zeta^{\rm R}$ ($\circ$). 
    (f) Normalized empirical probabilities for the residual forward $p(\zeta^\mathrm{F})$ and the residual reverse $p(\zeta^\mathrm{R})$ time series corresponding to the observation time $t=400~\rm{s}$. 
    (g) Irreversibility measure $\hat{\sigma}$  obtained as a function of observation time for control (purple circles) and pharamacologically treated conditions: Latrunculin-A (green diamonds) and (-)-Blebbistatin (yellow triangles). The estimated $\hat{\sigma}(t=400~\mathrm{s})$, differs significantly between the control and the Latrunculin-A or Blebbistatin-treated cells. This is supported by $t$-test results, yielding $p$-values of 0.047 and 0.029, respectively.
    For each cell, the total observation time was $400$~s at a sampling rate of $f_s=100$~Hz.}
    \label{fig:panel_3}
\end{figure}

Recently, a novel measure for the non-equilibrium nature of a system was introduced based on the quantification of time irreversibility offering a lower bound $\hat\sigma$ of entropy production $\sigma$ ~\cite{roldan2010estimating,roldan2012entropy}. 
This measure captures the divergence between the probabilities of experimentally observed  forward and reverse trajectories in terms of the Kullback-Leibler Divergence (KLD) of whitened residual signals. 
Following the approach in Ref.\cite{roldan2021quantifying}, we compute the residual time series $\zeta^\mathrm{F}(t)$ and $\zeta^\mathrm{R}(t)$ from measured time series, see Fig.\ref{fig:panel_3}\,(a)-(d) and Materials and Methods.
After whitening, the autocorrelation function of the residuals vanishes at all times, indicating that the whitened signal behaves as an independent and identically distributed process, see Fig.~\ref{fig:panel_3}\,(e).
In Fig.~\ref{fig:panel_3}\,(f), we depict the normalized probability distributions $p(\zeta^\mathrm{F})$ and $p(\zeta^\mathrm{R})$ of the obtained residual time series for the observation time $t=400~\rm{s}$ which are then used in the calculation of the KLD and $\hat\sigma$, see Eq.~\eqref{eq:HatSigma} and~\eqref{eq:KLD}  in the Materials and Methods.
The determined irreversibility measures $\hat{\sigma}$ calculated for different observation times averaged over all cells are plotted in Fig.~\ref{fig:panel_3}\,(g). 
Similar to Ref.~\cite{roldan2021quantifying}, we find that $\hat{\sigma}$ decreases with observation time and acquires an approximately constant value for $t\geq 300~\mathrm{s}$ for all conditions. 
Notably, the irreversibility measure $\hat{\sigma}$ decreases for both inhibition treatments including Latrunculin-treated cells. We conclude that upon turnover inhibition (Latrunculin-A treatment), irreversibility and effective temperature also show different trends. 

\section{Asymmetric spikes and thermal noise have an antagonistic effect on time irreversibility}
\begin{figure}
    \centering
    \includegraphics[scale=0.75]{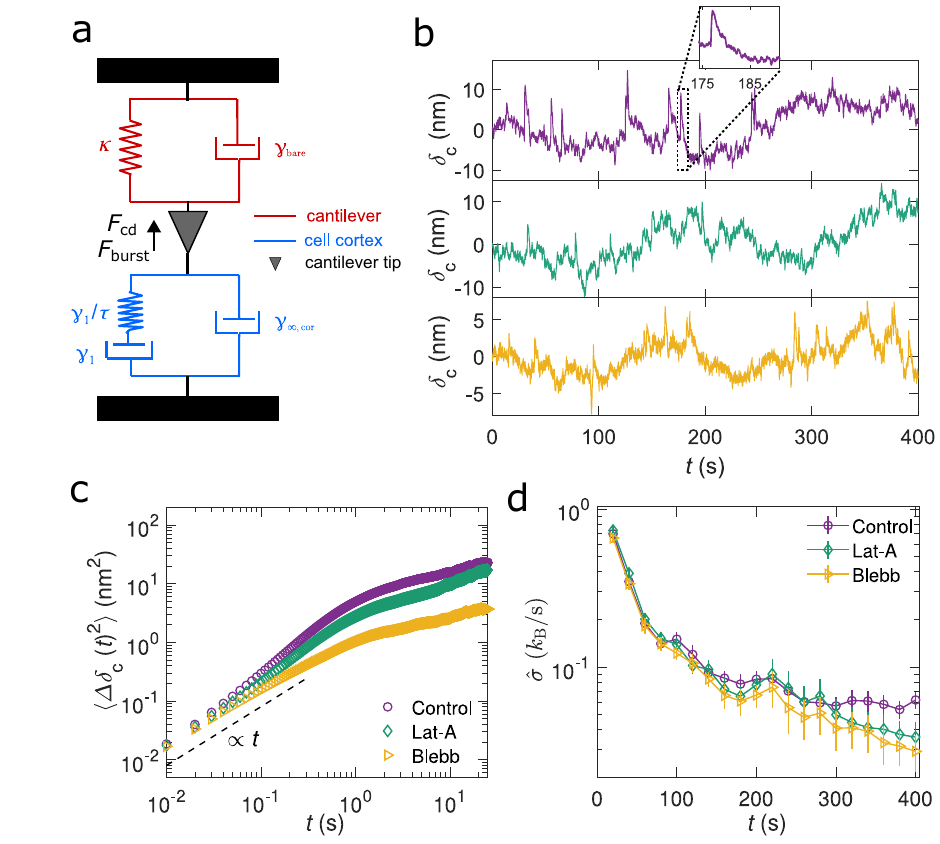}
    \caption{Minimal model conceptualizing trends in time irreversibility and thermal noise. (a) Schematic illustration of the mechanical environment of the cantilever tip when embedded in the cellular cortex. While the cantilever spring is captured by a Kelvin-Voigt model with cantilever spring constant $\kappa$ and friction coefficient $\gamma_{\rm bare}$, the cortex material is  modeled as viscoelastic Jeffrey's fluid. 
    (b) Representative numerically simulated trajectories based on the model described by Eqn.~\eqref{eq:numerical_model1}-\eqref{eq:numerical_model4}, corresponding to the control case (top), Latrunculin-treated cells (middle), and Blebbistatin-treated cells (bottom). (c) Mean squared displacements $\langle\Delta\delta_\mathrm{c}(t)^2\rangle$ obtained by averaging over 15 independent trajectories for parameters sets mimicking the control (purple), Latrunculin-treatment cells (middle), and Blebbistatin-treated cells (bottom). 
    (d) Entropy production rate $\sigma$ obtained using the Kullback-Leibler divergence approach from 15 independent numerically simulated trajectories of the AFM tip using the model described in (a).
    The simulated Langevin equation and corresponding parameters are given in Supplementary Section~\ref{sec:Sims}. 
    }
    \label{fig:panel_4}
\end{figure}

To better understand the inhibition-induced trend of the time irreversibility measure $\hat\sigma$ in our experimental data, we modeled the dynamics of the AFM tip embedded in the viscoelastic actomyosin cortex as an overdamped active Brownian probe subject to active forces. 
The mechanical response of the cortex is modeled by a viscoelastic Jeffrey's fluid, see Fig.~\ref{fig:panel_4}\,(a) and Supplementary Section~\ref{sec:Sims}. Further, the mechanical response of the cantilever is captured by a Kelvin-Voigt type response as characterized by experiments, see Fig.~\ref{fig:panel_4}\,(a).  
Activity is incorporated by scaling thermal forces by a factor $f_{\rm act} \geq 1$, see Supplementary Section~\ref{sec:Sims}. 
Additionally, we introduce active forces $F_{\rm cd}$ arising from the diffusion of the caging harmonic potential to account for the slow meandering of the AFM tip around its zero position observed in experimental trajectories. Lastly, we incorporate burst-like active forces, $F_{\rm burst}$, characterized by randomly occurring, short-lived force spikes - either symmetric downward or asymmetric upward - closely resembling the active force bursts observed in experiments (see Supplementary Section~\ref{sec:Sims} and Eqns.~\eqref{eq:numerical_model1}-\eqref{eq:numerical_model4}).


Exemplary simulation trajectories mimicking the three measurement conditions (purple: control, green: actin turnover inhibition, yellow: motor protein inhibition) are shown in Fig.~\ref{fig:panel_4}\,(b) and Fig.~\ref{supp_panel_8_new}. Parameters were adjusted such that all salient features of experimental trajectories were reproduced. In particular, force spike and dip amplitudes were reduced in the treatment conditions (green and yellow trajectories), consistent with experimental observations (see Fig.~\ref{fig:supp_panel_1_new}\,(c),(f) and Fig.~\ref{fig:supp_panel_3_new}).
Acknowledging the effect of cortex activity, thermal noise is increased by a frequency-dependent upscaling of thermal noise according to our measurements of effective temperature for the different conditions, see Supplementary Section~\ref{sec:Sims} and Fig.~\ref{fig:supp_panel_10_new}. Further, cortex mechanics parameters were varied according to rheological measurements under different experimental conditions (see Section~\ref{sec:Sims}). 
By computing 15 trajectories over $400$~s for each condition, we obtained the MSD and irreversibility measure $\hat\sigma$, see Fig.~\ref{fig:panel_4}\,(c),(d). Our results show that the MSD is lower in simulations mimicking inhibition treatments with a stronger reduction for motor protein inhibition, see Fig.~\ref{fig:panel_4}\,(c), similar to experimental results, see Fig.~\ref{fig:panel_2}\,(c). Further, $\hat\sigma$ is reduced for both turnover and motor inhibition compared to control, see Fig.~\ref{fig:panel_4}\,(d). 

Performing a systematic analysis, we identify the thermal noise scaling factor $f_{\rm act}$ and force spike amplitude $\mathcal{A}_u$ as key parameters regulating $\hat\sigma$.  In particular, we find that increasing the thermal noise amplitude via $f_{\rm act}$ decreases $\hat\sigma$ but increases the MSD, see Fig.~\ref{fig:supp_panel_9_new}\,(a),(d). Conversely, upscaling asymmetric force spikes through the amplitude $\mathcal{A}_u$ increases both $\hat\sigma$ and the MSD, see Fig.~\ref{fig:supp_panel_9_new}\,(b),(e).
This analysis suggests that for turnover inhibition, reduction of $\hat\sigma$ results from diminished force spike amplitudes. For motor inhibition, reduction of time irreversibility $\hat\sigma$ is primarily due to a significant drop in force spike amplitude, despite lower thermal noise.
Interestingly, we see that when both force spike amplitude $\mathcal{A}_u$ and thermal noise amplitude $f_{\rm act}$ are upscaled jointly, $\hat\sigma$ shows still a net increase, see Fig.~\ref{fig:supp_panel_9_new}\,(c) and (f). Therefore, a coupled up-regulation of all active force contributions leads to enhanced time irreversibility. 

Our theoretical analysis suggests that \textit{a priori}, there is no reason for why time irreversibility and effective temperature should in general exhibit similar trends upon perturbations as opposite trends of these two quantities can be induced through e.g. a change in the thermal noise amplitude. However, such a phenomenon might be avoided in nature by a coupled up- or downscaling of thermal noise and time-asymmetric deflection spikes in living systems. 

In the case of the actin cortex, this coupling may arise through the following plausible mechanism: increased thermal-like noise raises the probability of tension-gated calcium channels to open, giving rise to a short-term calcium spike and concerted myosin motor activation \cite{vicente-manzanares_non-muscle_2009}. The induced force spike equilibrates over time due to a gradual myosin deactivation and restoration of preceding myosin activity levels \cite{kovacs_functional_2003}.

In general, we propose that the coupling between the strength of thermal noise and time asymmetric ‘excitation-restoration processes’ may represent a prevalent mechanism in biological systems, whereby the magnitude of thermal-like noise regulates the frequency and amplitude of time-asymmetric deflection spikes for instance triggered by  i) threshold-gated excitable dynamics of biological signaling, see e.g. \cite{al_beattie_criticality_2024, suzuki_dendritic_2017, zhou_bifurcation_2019}, or ii) material failure and subsequent regulated repair processes \cite{charras2006reassembly, hui_coordinated_2023}.

\section{discussion}
In this study, we investigated active position fluctuations $\delta_\mathrm{c}=\Delta F/\kappa$ of a soft AFM cantilever tip embedded within the cortex of mitotic human cells. Our work represents the first experimental realization of AFM-based measurements of active fluctuations. Unlike previous studies in interphase cells, see e.g. \cite{schlosser_force_2015, ahmed_active_2018, turlier2016equilibrium, hurst_intracellular_2021}, we observed distinct active force bursts in the AFM readout — corresponding to abrupt upward or downward displacements of the AFM probe tip. We conjecture that these bursts arise from pulses of synchronized myosin motor activation (position spikes) and sudden cortical material failure (position dips). 
Active bursts significantly enhanced the MSD of the cantilever tip leading to a transition from diffusive to super-diffusive behavior indicating the presence of active fluctuations in addition to passive thermal fluctuations. Quantifying the power spectral density and the response function of the cantilever tip position, we find that the effective temperature of cantilever fluctuations is significantly increased by an activity factor of up to $\approx 100$ for timescales $\gtrsim 1$~s in agreement with earlier results from tweezer measurements \cite{hurst_intracellular_2021}. 
We observed that force spikes showed an asymmetric shape with a rapid rise and a slower exponential relaxation with a characteristic time of \( \tau \approx 1.5~\mathrm{s} \).  Correspondingly, we find a non-vanishing time reversal asymmetry measure $\hat\sigma$ derived from the KLD between the whitened forward and time-reversed signal. 

To probe the influence of biological activity on the system's fluctuations, we inhibited two active processes in the actin cortex – i) the dynamic turnover of the cortical actin matrix and ii) the activity of cortex-associated motor proteins. While turnover inhibition resulted in only a minor change in the effective temperature of fluctuations, motor protein inhibition led to a dramatic decrease, underscoring the crucial role of motor proteins in driving active fluctuations. 
Counterintuitively, despite exhibiting similar values of \( T_\mathrm{eff}(f) \) as the control, turnover-inhibited cells displayed a lower entropy production rate \( \sigma \), as inferred from the Harada-Sasa equality. This can be attributed to the more solid-like elastic nature of the cortex in turnover-inhibited cells, leading to reduced energy dissipation for a given noise level.  Therefore, consistently, we observe that entropy production decreases for both types of activity reduction. However, this trend is not reflected in both cases by a corresponding change in effective temperature.

Finally, we find a reduced time irreversibility \( \hat{\sigma} \) in both types of activity-inhibited cells. Thus, the changes of the time irreversibility measure $\hat \sigma$ agree qualitatively with the changes of entropy production $\sigma$. 
To rationalize these trends in time irreversibility, we reproduced experimental tip trajectories in simulations of an overdamped Brownian probe embedded in a non-Markovian viscoelastic medium subjected to active force bursts and upscaled Brownian motion. Our simulations indicate that time irreversibility in activity-inhibited cells is reduced due to a decrease in force spike amplitudes. Interestingly, our simulations also show that a mere upscaling of thermal noise (and correspondingly of effective temperature)  leads to a reduction of the time irreversibility. 

In conclusion, our study demonstrates that different non-equilibrium measures such as effective temperature \( T_\mathrm{eff} \), entropy production $\sigma$ and time irreversibility \( \hat{\sigma} \) can exhibit contrasting behaviors in active living systems, depending on the nature of the active processes driving the motion of embedded probes. Notably, our findings challenge the validity of effective temperature as a proxy for the distance from thermal equilibrium.
While we find similar trends for  $\sigma$ and $\hat{\sigma}$ in our experimental measurements, our simulations suggest that perturbation types exist that might change these two quantities in opposite ways. 
It remains an intriguing open question whether a fundamental law governs the coupling of entropy production rate and time irreversibility in living systems.

\section{Materials and Methods}
\subsection{Cell Culture}
We cultured the HeLa Kyoto cells in the growth medium DMEM, ($\#31966-021$, Life Technologies) in combination with $10\,\%$~v/v fetal bovine serum (FBS, $\#10270106$, Life Technologies) and $1\,\%$~v/v penicillin/streptomycin ($\#15140122$, Life Technologies) at $T=37^\circ\mathrm{C}$ and $5\,\%$ $\mathrm{CO_2}$. The splitting was performed every 2 to 3 days upon reaching a confluency of $60–80~\%$. 
Around $10,000$ cells were seeded a day prior to the experiment at the center of a $3.5~\mathrm{cm}$ glass petri-dish. The seeding area was confined within a silicon chamber with a media volume of $250~\mathrm{\mu L}$ ensuring precise spatial localization of the cells in the center of the dish. On the measurement day, the silicon chamber was removed, and the growth medium was replaced with $2~\mathrm{mL}$ $\mathrm{CO_2}$-independent DMEM ($\#\mathrm{12800-017}$, Invitrogen) containing $4~\mathrm{mM}~\mathrm{NaHCO_3}$, buffered with $20~\mathrm{mM}~\mathrm{HEPES/NaOH}$ at $\mathrm{pH}=7.2$, and enriched with $10\%$ FBS. Additionally, the medium was supplemented with $2~\mu\mathrm{M}$ S-trityl-l-cysteine (STC, $\#\mathrm{164739}$, Sigma) which inhibits kinesin Eg\,5 arresting cells in mitosis. Cells were then kept in an incubator maintained at $T=37^\circ\mathrm{C}$ without $\mathrm{CO_2}$ for 2 to 3 hours until mitotic cells were visibly enriched.

For measurement of drug-treated cells, pharmacological agents $(-)-$Blebbistatin ($10~\mathrm{\mu M}$, $\#\mathrm{B0560-1MG}$, Sigma) or Latrunculin-A ($25~\mathrm{nM}$, $\#\mathrm{AG-CN2-0027}$, AdipoGen Life Sciences) were added to the medium at least $15~\mathrm{min}$ prior to measurements. 

\subsection{Atomic Force Microscopy Measurements}
The experiments were performed with a Zeiss Axiovert inverted light microscope combined with an AFM Nanowizard 4XP (JPK Instruments) using a $20\times$/ 0.8 air objective. 
The petri dish containing cells were maintained at $37^{\circ}\mathrm{C}$ using a petri dish heater (JPK Instruments). All measurements were performed using triangular MLCT-D cantilevers (Bruker) with nominal spring constant $\kappa=0.03~\mathrm{N/m}$. On each day of the measurement, the spring constant $\kappa$ of the cantilever was separately determined using the built-in software (JPK Instruments) which exploits the power spectral density of the thermal fluctuations.

\subsubsection*{Measuring Active Fluctuations and Rheology}
First, a cell in mitotic arrest was chosen. Then an approach was performed on top of this cell with the tip touching approximately the center of the projected cell area. After cantilever retraction, the cantilever was extended with a speed of $2~\mathrm{\mu m/s}$ up to a peak force of $1~\mathrm{nN}$. Subsequently, the piezo-controlled cantilever hinge height was kept constant for a time interval of $500~\rm{s}$. After the initial force peak, the tip deflection was allowed to equilibrate  for $100~\mathrm{s}$. Then, the measurement was performed for further $400~$s where the time $t$, hinge height $h$ and the tip deflection $F$ was recorded for later analysis of tip height fluctuations $\delta_\mathrm{c}=F/\kappa$. Depending on the measurement, the sampling frequency was kept between $100-1000~\mathrm{Hz}$. 

For rheology measurements, $h$ was varied sinusoidally across a range of frequencies from $0.02~\mathrm{Hz}$ to $10~\mathrm{Hz}$, with an amplitude of $200~\mathrm{nm}$. Figure~\ref{fig:supp_panel_4_new} presents a typical time evolution of the oscillating cantilever hinge height $h$ and the corresponding tip position oscillations $\delta_\mathrm{c}$ for frequencies between $0.05~\mathrm{Hz}$ and $10~\mathrm{Hz}$.

\subsubsection*{Mean Squared Displacements}
We first quantified the active nature of AFM tip fluctuations embedded within the cortex by the mean squared displacement $\langle \Delta \delta_\mathrm{c} (t)^2\rangle=\langle[\delta_\mathrm{c}(t+t_0)-\delta_\mathrm{c}(t_0)]^2\rangle$ which are plotted in Fig.~\ref{fig:panel_2}\,(a). In order to calculate $\langle \Delta \delta_\mathrm{c} (t)^2\rangle$, a time average over $t_0$ is first computed for a given trajectory and then ensemble average is performed over the different realizations on cells $N$ for each case.

\subsubsection*{Microrheology}
Following the Refs.~\cite{AFM_alcaraz2003microrheology, AFM_rother2014},
we compute the frequency-dependent complex elastic shear modulus $G_{\rm eff}^*(f)$ of the cortical layer from the externally imposed sinusoidal oscillations of different frequencies of the tip motion, see Fig.~\ref{fig:supp_panel_6_new}, using 
\begin{equation}
   G^*_{\rm eff}(f)=\frac{\sqrt{2}}{6\tan(\beta)\delta_0}\frac{\hat{F}(f)}{\hat{\delta}_h(f)}  \exp(i\Delta\varphi)
\end{equation}
where $\hat{F}(f)$ and $\hat{\delta}_h(f)$ are the fitted amplitudes of the sinusoidal force signal and the indentation depth $\delta_h(t)=-\left( h(t)+\delta_\mathrm{c}(t)\right)$ with $h(t)$ the piezo height of the AFM. Here, $\beta=17.5^\circ$ is the half-angle to edge of the pyramidal tip of the MLCT cantilever and $\delta_0$ is the average indentation depth within the cortex as determined from our measurement separately on each cell. We find that $\delta_0$ typically adopts values of $\approx 1~\mathrm{\mu m}$. Beyond contributions from viscoelastic stresses, measured force amplitudes are expected to show contributions that stem from active stresses coupling to geometrical changes, see \cite{fischer-friedrich2018}. 

Given the molecular turnover in the cortex, we can infer that active stress contributions dominate the force signal at small frequencies \cite{fischer-friedrich2018}. Therefore, we estimate the activity-corrected complex elastic shear modulus $G^*(f)$ by   $G^*(f)=G^*_{\rm eff}(f)-G_{\rm eff}'(0.02~{\rm Hz})$, see Fig.~\ref{fig:supp_panel_5_new}.

\subsubsection*{Fluctuation Dissipation Theorem}
The FDT provides a direct link between mechanical properties captured by the response function $\chi(f)$ and the strength of fluctuations in the linear regime. Its violation indisputably confirms the deviations from equilibrium. Mathematically, it relates the power spectral density $C(f)$ of the tip fluctuations $\delta_\mathrm{c}(t)$ to the imaginary part of the response function $\chi''(f)$ as
\begin{equation}
\frac{2\pi C(f)}{2k_\mathrm{B}T}=\chi''(f)
\label{eq:fdt}
\end{equation} 
In our experiments, $C(f)$ was obtained from the Fast Fourier Transformation of the measured tip fluctuations $\delta_\mathrm{c}(t)$ using MATLAB as
\begin{equation}
    \tilde{\delta}(f)=\rm{FFT}(\delta_\mathrm{c}(t))\hspace{1cm}
    C(f)=\tilde{\delta}(f)\tilde{\delta}^*(f)/f_{s} p    
\end{equation}
where $\tilde{\delta}^*(f)$ is the complex conjugate of $\tilde{\delta}(f)$, $p$ is the number of measurement points in $\delta_\mathrm{c}$, and $f_s$ is the sampling rate. While the imaginary part of the response function $\chi''(f)$ was determined by performing sinusoidal indentations in the cell cortex at various frequencies $f$ with amplitude $200~\mathrm{nm}$ and measuring the resultant linear response $\chi(f)=\hat{\delta}_h(f)/\kappa \hat{h}(f)$. We compute the effective temperature as $T_\mathrm{eff}(f)=\pi C(f)/k_{\mathrm{B}} T\chi''(f)$ is computed from the experimental data, we performed binning of the response measured from $C(f)$ over a frequency window with width $\epsilon=0.01~\mathrm{Hz}$, \textit{i.e.}, $f\pm\epsilon/2$. 

\subsection{Calculation of the Kullback-Leibler Divergence}
Following Rold\'an {\it et al.} \cite{roldan2021quantifying}, we determine the irreversibility measure $\hat\sigma$ by calculating the Kullback-Leibler Divergence (KLD) of the measured stochastic time series $\delta_\mathrm{c}(t)$ of the AFM tip position embedded within the cortex for both the treated and untreated cases. 
For this purpose, we obtained from the experimentally measured time series $\delta_\mathrm{c}^\mathrm{F}(t)$, the time-reversed series $\delta_\mathrm{c}^\mathrm{R}(t)$ by inverting the time index, as demonstrated in Fig.~\ref{fig:panel_3}\,(a) and (b). An autoregression model AR$(m)$ of order $m=25$ is then obtained by fitting the reverse time series using the MATLAB “arfit” package developed by Schneider and Neumaier~\cite{schneider2001algorithm}. Using the obtained AR$(25)$ model, the residual time series for both the forward and the corresponding reverse time series are computed as
\begin{equation}
 \xi^\mathrm{F,R} =
  \begin{cases}
    \delta_{\mathrm{c}}^i & \text{if} \; i \leq m \\
    \delta_{\mathrm{c}}^i - \sum_{j=1}^{m} a_j \delta_{\mathrm{c}}^{i-j} & \text{if} \; i \geq m
  \end{cases}
  \label{eq:whitening}
\end{equation}
where $\delta_\mathrm{c}^i$ is the $i^{th}$ element of the forward/reverse time series $\delta_\mathrm{c}^\mathrm{F,R}(t)$ while $a_j$ is the $j^{th}$ coefficient of the AR model, see Fig.~\ref{fig:panel_3}\,(c) and (d). 
The residual time series corresponding to the forward and reverse trajectories are plotted in Fig.~\ref{fig:panel_3}\,(c) and (d). 
This process is  called whitening, as it removes the time correlations present in the time series, rendering the resultant time series $\xi_c^\mathrm{F,R}(t)$ as an independent and identically distributed (i.i.d.) process. We choose $m=25$ for analysis as it effectively removes all the time-correlations present in the measured trajectories. This is demonstrated by the autocorrelation functions for both the residual forward $\zeta^\mathrm{F}(t)$ and residual reverse $\zeta^\mathrm{R}(t)$ time series in Fig.~\ref{fig:panel_3}\,(e).
For comparison, we also show the autocorrelation functions and consequently the KLD for various values of $m$ in the supplementary Fig.~\ref{fig:supp_panel_7_new}. 
As shown in the inset Figs.~\ref{fig:supp_panel_7_new}\,(a)-(c), for $m<25$ the autocorrelation functions of the residuals do not vanish entirely but remain finite, particularly for times $t\lesssim 1~\mathrm{s}$. In Fig.~\ref{fig:panel_3}\,(f), we depict the normalized probabilities $p(\zeta^\mathrm{F})$ and $p(\zeta^\mathrm{R})$ which were computed considering 50 bins spanning from the minimum to the maximum values of the residual series. 
The irreversibility measure $\hat{\sigma}$ is calculated as  
\begin{equation}
\label{eq:HatSigma}
    \hat{\sigma} \geq k_\mathrm{B} f_s D[p(\zeta^\mathrm{F})||p(\zeta^\mathrm{R})]
\end{equation}
where  
\begin{equation}
\label{eq:KLD}
    D = \gamma \sum_i p(\zeta_i^\mathrm{F}) \ln \left( \frac{p(\zeta_i^\mathrm{F})}{p(\zeta_i^\mathrm{R})} \right)
\end{equation}
is the KLD. The prefactor $\gamma$ accounts for the statistical bias of the KLD estimate \cite{roldan2021quantifying}. It is obtained using the Kolmogorov-Smirnov (KS) test under the null hypothesis $H_0: p(\zeta^\mathrm{F}) = p(\zeta^\mathrm{R})$, which yields a p-value $p_{KS}$. The prefactor $\gamma$ is defined as $\gamma = 1 - p_{KS}$.

Corresponding results for $\hat \sigma$ are found in Fig.~\ref{fig:panel_3}\,(g). 
Remarkably, determined \( \hat{\sigma} \) values were comparable to those observed in the cortex of starfish oocytes \cite{tan2021scale}. 

\section*{Acknowledgments}
We thank Nir Gov and Paul Janmey for their helpful feedback on the project. 
Furthermore, EFF thanks the Deutsche Forschungsgemeinschaft (DFG, German Research Foundation) for financial support through the Heisenberg program, project number 495224622 (FI 2260/8-1) and the grant FI 2260/5-2. 
In addition, the authors thank the PoL Light Microscopy Facility for their excellent support.

\section*{Competing Interests}
The authors declare no competing interests.

\section*{Data availability statement} 
All data generated or analyzed during this study are available from https://doi.org/10.25532/OPARA-792. Generated code for analysis and simulations can be downloaded from https://gitlab.com/polffgroup/corticalfluctuations.

\section*{Author Contributions}
N.N. performed the experiments and data analysis. N.N. and E.F.-F. designed the experiments. N.N. and E.F.-F. wrote the manuscript.


\newpage
\pagebreak
\setcounter{figure}{0}
\setcounter{section}{0}
\setcounter{subsection}{0}
\begin{suppenv}
\section*{Supplementary Material}

\subsection{Spatiotemporal characterization of force bursts}
\label{sec:trajectory_char}
In order to understand how the main features of the measured stochastic trajectories of $\delta_\mathrm{c}$ influence the irreversibility measure $\hat{\sigma}$, we analyzed the observed key features of the stochastic trajectories, specifically the upward and downward-directed active force bursts. These bursts were characterized by their amplitude, the characteristic relaxation time (which could be reliably fitted with an exponential function only for upward bursts), and the frequency of such events within a time window of $400~\mathrm{s}$. The amplitudes and number of the peaks were determined applying the MATLAB inbuilt function \textit{findpeaks} to the time series of $\delta_\mathrm{c}$ with additional constraints such as minimum peak height and width and the gradient with which a peak rises which defines the prominence of the peak. 
An offset correction to each peak is performed to obtain its actual height. This was done by finding the minimum y-value in the trajectory within a time interval of $2~\mathrm{s}$ preceding the peak location. 
Finally, the actual peak amplitude is the difference between measured peak value and the minimum y-value providing $A_u$. 
By fitting $\delta_\mathrm{c}$ with an exponential decay $C \exp(-t/\tau_u)$ in a time interval of $2.2$~s after the peak, we determined a corresponding relaxation time scale $\tau_u$, see Fig.~\ref{fig:supp_panel_1_new}\,(a) and (b).
To determine amplitudes of force dips $A_d$, the peak-analysis procedure was repeated for $-\delta_\mathrm{c}$. 
The obtained results are displayed as boxplots in Fig.~\ref{fig:supp_panel_1_new}.
Notably, the amplitudes of both upward and downward active force bursts exhibited significant differences between the control and treated cases. 
In particular, both $A_u$ and $A_d$ are the highest in the control cells, reduced in the Latrunculin-treated cells, and lowest in the Blebbistatin-treated cells. 
This is consistent with the magnitude differences of observed MSDs (Fig.~\ref{fig:panel_2}\,(a),  main text) and the corresponding entropy production rates $\sigma$ obtained from FDT violation. 

In contrast to the amplitudes, the characteristic relaxation timescales $\tau_u$ of the upward active force bursts and the number of force bursts measured over $400~\mathrm{s}$ remained rather similar for all the cases. Although the number of downward bursts in the Blebbistatin-treated cells showed a significant increase (with $p < 0.05$), this change did not correspond to a significant effect on the measured $\hat{\sigma}$. This suggests that the amplitude of the bursts had a more dominant influence than the number of bursts.

Additionally, we characterized the temporal asymmetry of the observed upward and downward force bursts, see insets of Fig.~\ref{fig:panel_1}\,(c) main text. To achieve this, we measured the time intervals between the left half-maximum and the peak position of each burst, denoted as $\tau_u^l$ and $\tau_d^l$ for the upward and downward peaks, respectively. Similarly, the time intervals between the peak and the right half-maximum were labeled as $\tau_u^r$ and $\tau_d^r$.
To facilitate comparison, we plotted the corresponding left and right halves against each other for both the control and treated cases, as shown in Fig.~\ref{fig:supp_panel_1_new}. Interestingly, while the downward peaks also displayed temporal asymmetry, it was less pronounced compared to the upward peaks. 
To quantify the relative temporal asymmetries of force bursts for control and treated cases, we plotted boxplots of the ratios $\tau_u^r/\tau_u^l$ and $\tau_d^r/\tau_d^l$ for each condition. We find that upwards bursts tend to be more asymmetric then force dips, compare Fig.~\ref{fig:supp_panel_1_new}(k) and (o). Further, for both bursts and dips, asymmetries were more pronounced in Latrunculin-treated cells than in the control group. 

\begin{figure}[H]
    \centering
    \includegraphics[scale=0.6]{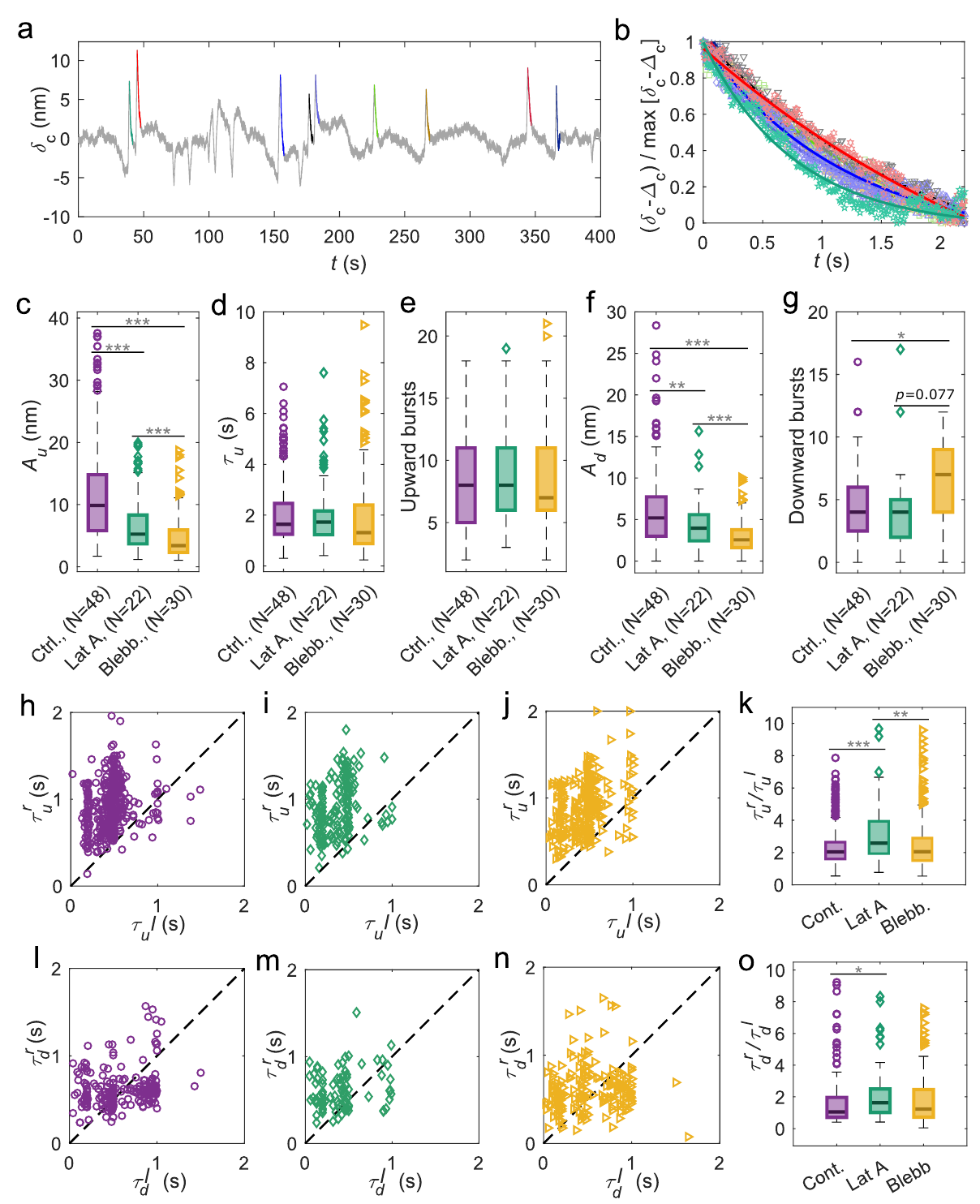}
    \caption{(a) Representative trajectory of an AFM tip inserted in the cortex of an untreated cell, highlighting the relaxation process following upward force bursts. (b) Normalized tip position coordinate $(\delta_\mathrm{c}-\Delta_\mathrm{c}) / \mathrm{max} [\delta_\mathrm{c}-\Delta_\mathrm{c}]$ following each upward force burst, where $\delta_\mathrm{c}=\delta_\mathrm{c}(t_{\rm peak}+2.2~{\rm s})$. The solid black curves represent the corresponding exponential fit. (c)-(g) Boxplots summarizing key characteristics of AFM tip trajectories within the mitotic cortex of untreated cells (purple) and cells treated with Latrunculin-A (green) or Blebbistatin (yellow). (c) Amplitudes of upward ($A_u$), outward-directed) active force bursts. (d) Characteristic relaxation time ($\tau_u$) following an upward force burst, determined by fitting a single exponential decay function, as illustrated in (b). (e) Number of upward bursts recorded over a 400 s measurement period per cell. (f) Amplitudes of downward ($A_d$, inward-directed) active force bursts. (g) Number of downward bursts recorded over 400 s per cell. In (c), (d), and (f), each data point represents a single force burst, whereas in (e) and (g), each data point corresponds to an individual cell. Boxplots were generated using data from $N=48$ control cells, $N=22$ Latrunculin-A-treated cells, and $N=30$ Blebbistatin-treated cells. The total number of analyzed upward bursts was 219, 92, and 194, respectively, while the number of downward bursts was 414, 174, and 270. (h)-(j) Comparison of the timescales for the left and right sides of upward active force bursts along the measured forward time trajectories for the untreated (purple circles), Latrunculin-treated (green diamonds) and the Blebbistatin-treated cells (yellow triangles). The times $\tau_u^l$ and $\tau_u^r$ represent the time intervals between the left half-maximum and the peak, and between the peak and the right half-maximum, respectively, along a forward time trajectory. (k) Boxplot depicting the ratio of the time interval between the right and left half-maximum from upward peaks. (l)-(n) Comparison of the timescales for the left $\tau_d^l$ and right $\tau_d^r$ side the downward active force bursts measured along the forward time trajectories. (o) Boxplot showing the ratio of the time interval between the right and left half-maximum from downward peaks. The box plot is obtained by analyzing the following number of measured cells: $N=48$ for control, $N=22$ for Latrunculin-treated and $N=30$ for Blebbistatin-treated cells. The corresponding number of total force bursts analyzed were $219$, $92$ and $194$, respectively. Statistical significance ($***~~p<0.001$) for each boxplot was assessed using a two-sample $t-$test.}
    \label{fig:supp_panel_1_new}
\end{figure}

\begin{figure}[H]
    \centering
    \includegraphics[scale=0.6]{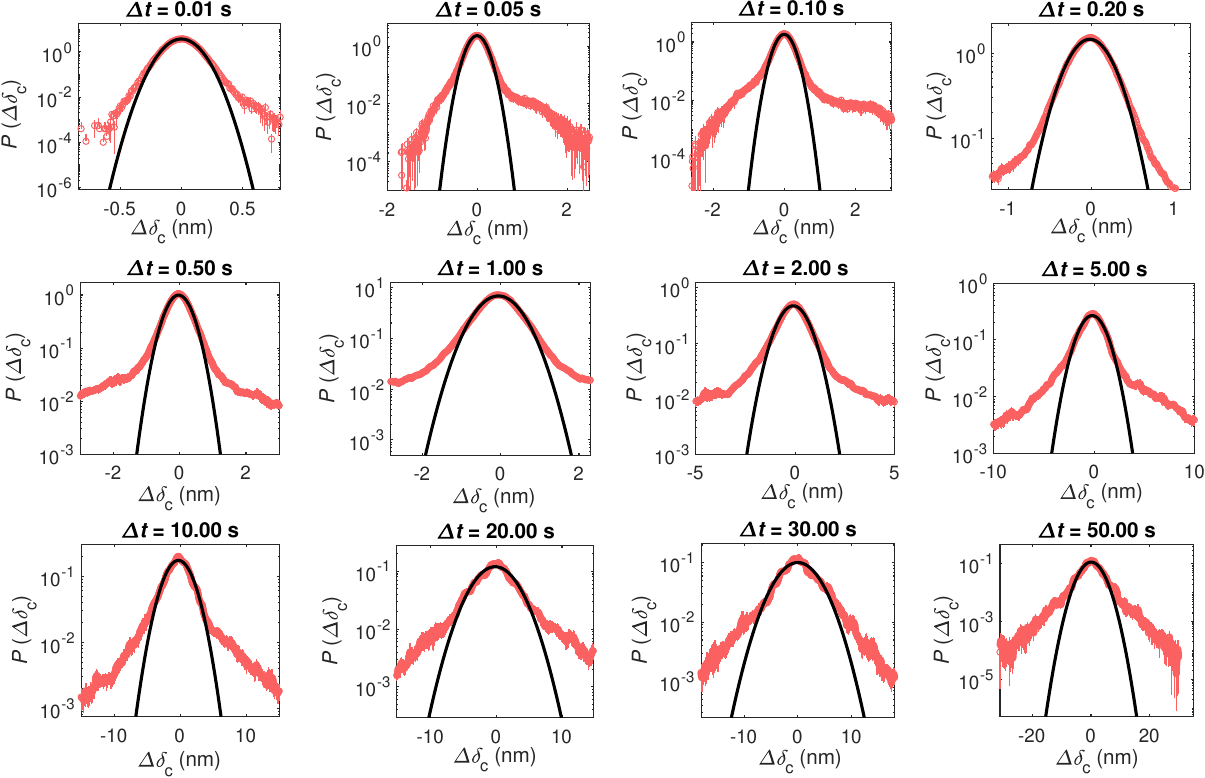}
    \caption{Normalized probability distribution function $P(\Delta\delta_\mathrm{c})$ of AFM tip displacements at different time lags $\Delta t$ labelled on each plot, averaged over 10 trajectories for control cells. The solid black curve in each case represents the corresponding Gaussian fit.}
    \label{fig:supp_panel_2_new}
\end{figure}

\begin{figure}[H]
    \centering
    \includegraphics[scale=0.55]{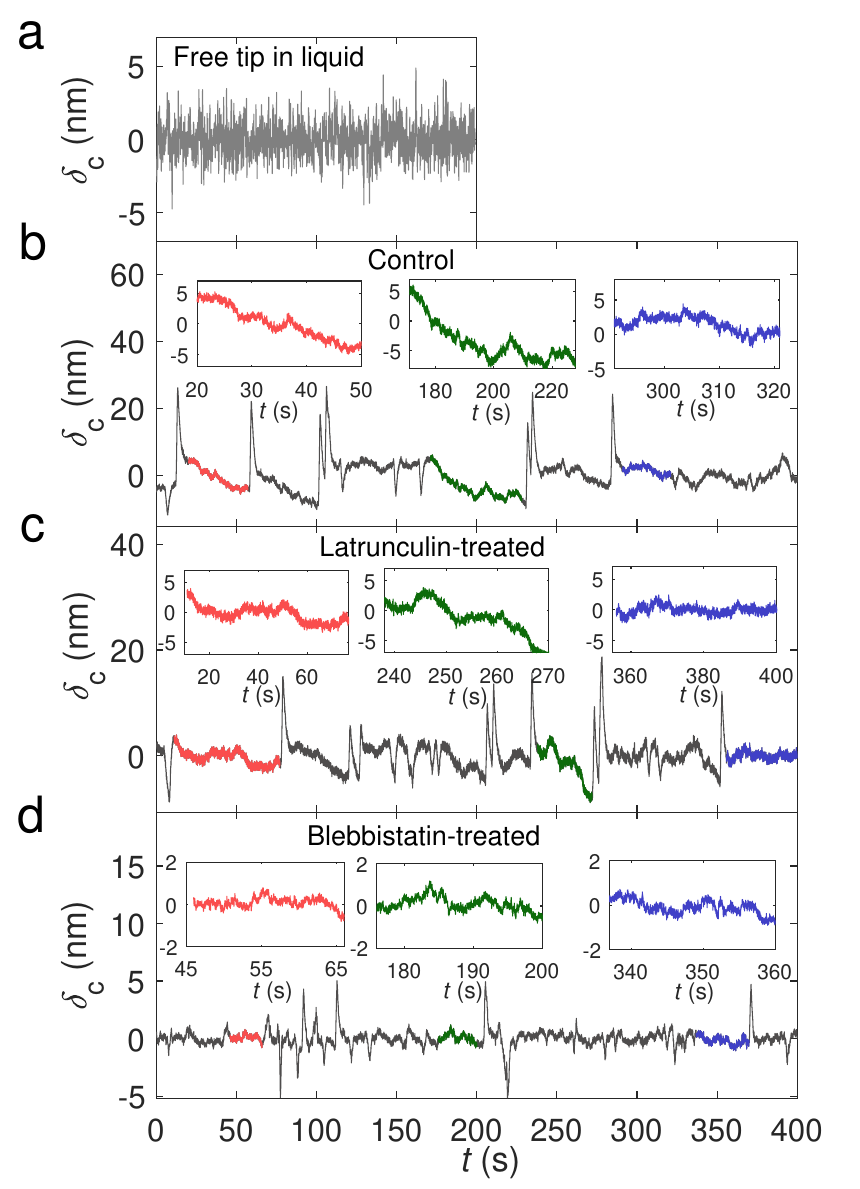}
    \caption{Representative trajectories of an AFM tip under different conditions. (a) Freely fluctuating in cell culture medium. (b) Embedded within mitotically arrested, untreated HeLa cells. (c) Embedded in a Latrunculin-treated cell ($25~\mathrm{nM}$). (d) Inserted into a Blebbistatin-treated cell ($10~\mathrm{\mu M}$). Insets in (b)-(d) provide close-up views of selected trajectory segments showing the high-frequency fluctuations.}
    \label{fig:supp_panel_3_new}
\end{figure}

\begin{figure}[H]
    \centering
    \includegraphics[scale=0.53]{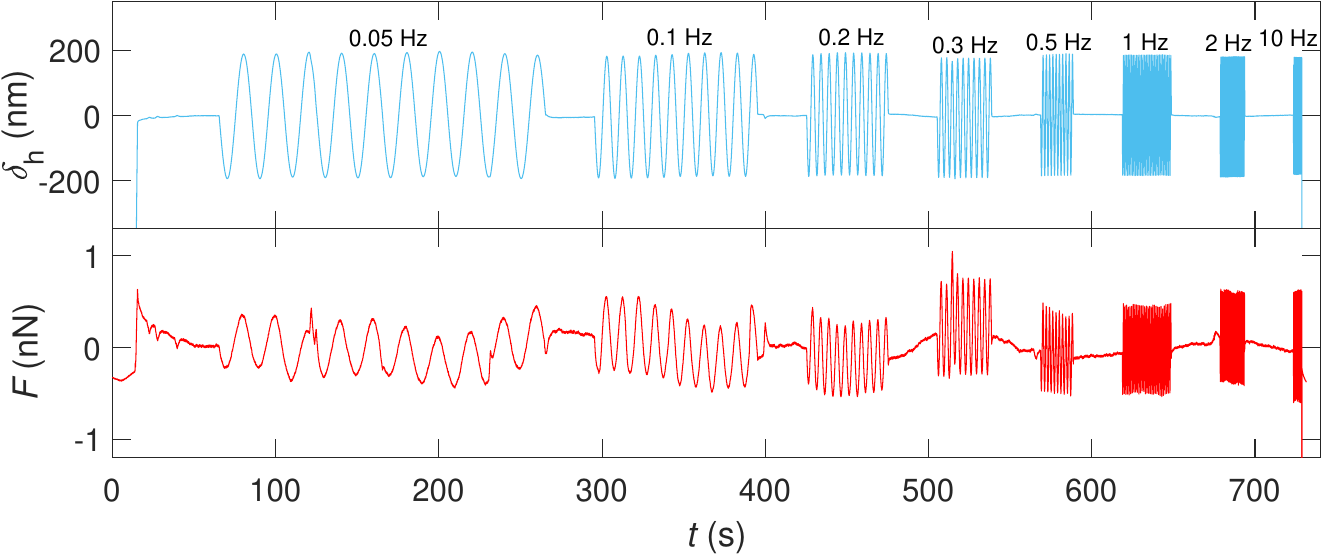}
    \caption{(Upper panel) Oscillatory measurements for rheological analysis. Upper panel: Time evolution of the tip height $\delta_h$ of the AFM cantilever oscillating at different frequencies while embedded in the cortical region of a mitotic cell.  
Lower panel: Time evolution of the measured force $F$ of the AFM cantilever in response to the oscillatory motion of the hinge height.} 
    \label{fig:supp_panel_4_new}
\end{figure}

\begin{figure}[H]
    \centering
    \includegraphics[scale=0.6]{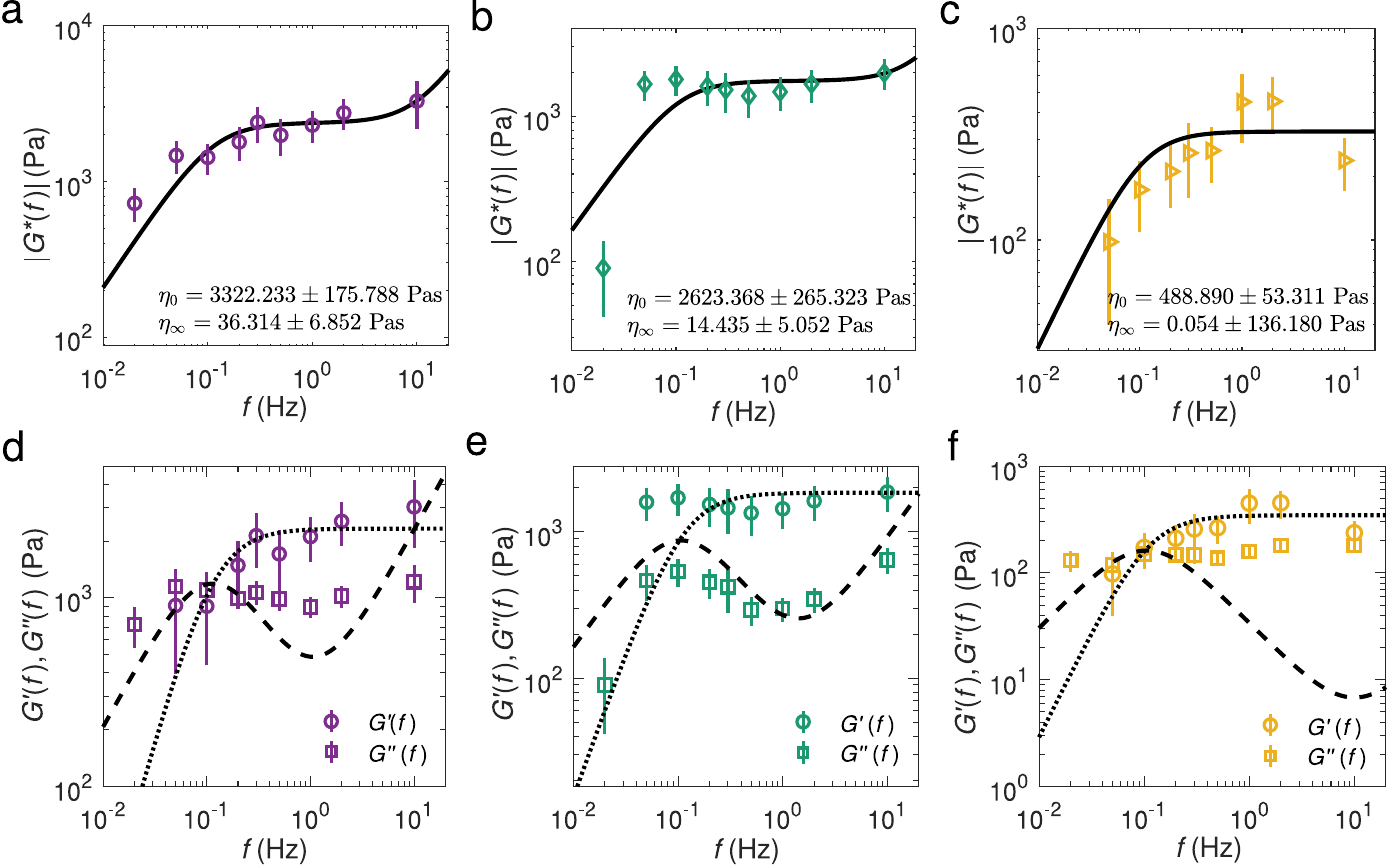}
    \caption{(a)-(c) Frequency dependent corrected complex shear modulus $G^*(f)$ measured by subtracting the lowest frequency \textit{i.e.}, $0.02~\mathrm{Hz}$ contribution from the real part of $G_\mathrm{eff}^*(f)$ in response to a sinusoidal perturbation of an embedded AFM tip into the cortex HeLa cells with an amplitude $200~\mathrm{nm}$ under (a) control (b) Latrunculin-treated ($25~\mathrm{n M}$) and (c) Blebbistatin treated ($10~\mathrm{\mu M}$) cells. The solid black curve is the Jeffreys model fitting corresponding to each (a)-(c). (d)-(f) The elastic $G'(f)$ and the loss $G''(f)$ moduli which correspond to the real and the imaginary part of the complex shear modulus $G^*(f)$, respectively for (d) control (e)  Latrunculin-treated ($25~\mathrm{n M}$) and (f) Blebbistatin treated ($10~\mathrm{\mu M}$) cells. The dotted and dashed curves are the corresponding Jeffreys model predication with rheological parameters $\eta_0$, $\eta_\infty$ obtained from fitting the corresponding complex shear modulus $G^*(f)$ and $\tau=1.5~\mathrm{s}$.}
    \label{fig:supp_panel_5_new}
\end{figure}

\begin{figure}[H]
    \centering
    \includegraphics[scale=0.6]{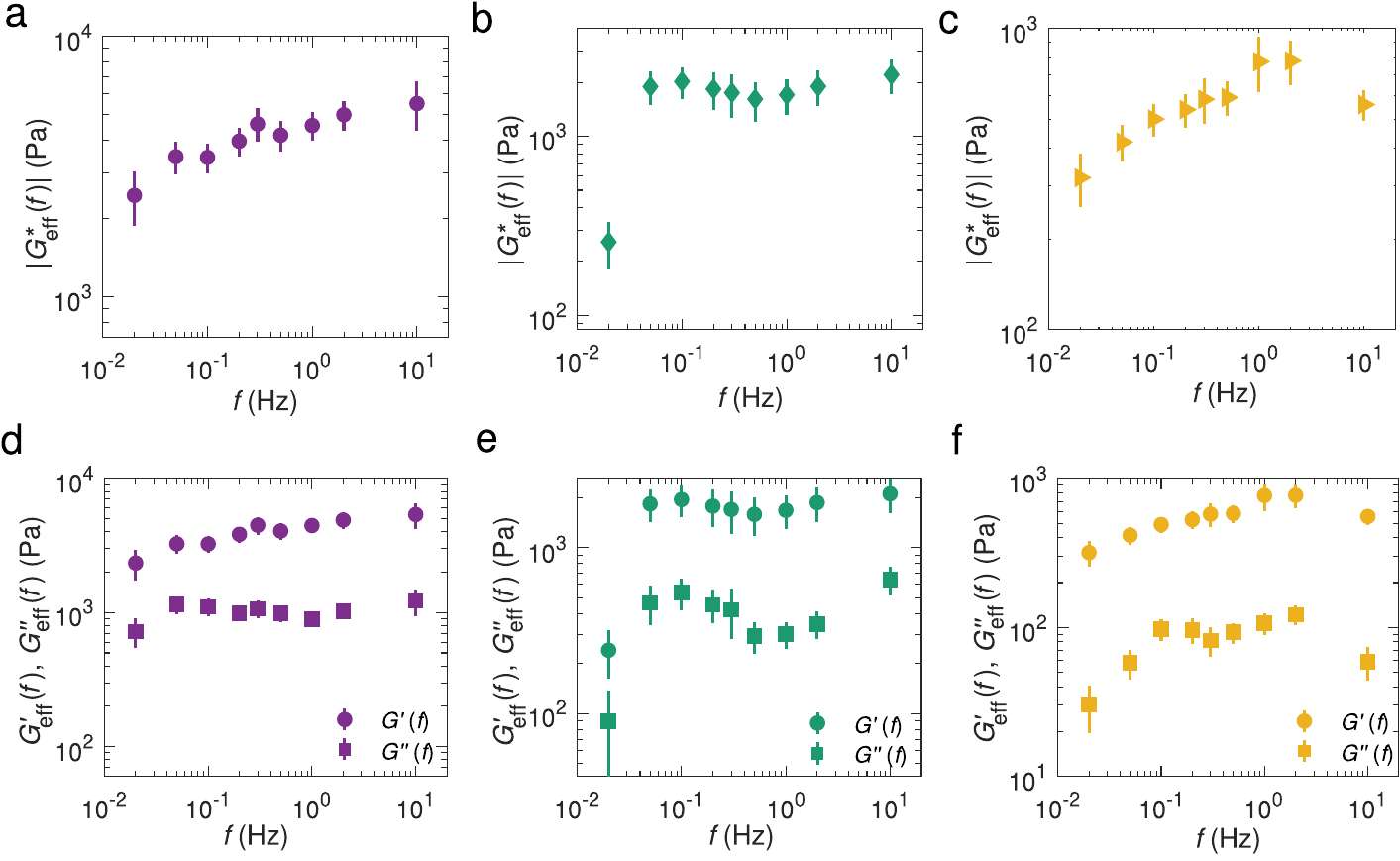}
    \caption{(a)-(c) Effective complex shear modulus \( |G_\mathrm{eff}^*(f)| \) of the cell cortex as a function of frequency, measured by sinusoidally oscillating the AFM tip in vertical direction with an amplitude of \( \approx 200~\mathrm{nm} \), under the following conditions: (a) control, (b) Latrunculin-A treatment ($25~\mathrm{nM}$), and (c) Blebbistatin treatment ($ 10~\mathrm{\mu M}$).
 The solid black curve is the Jeffreys' model fitting corresponding to each (a)-(c). 
 (d)–(f) Storage modulus $G'_\mathrm{eff}(f)$ and loss modulus $G''_\mathrm{eff}(f)$, representing the real and imaginary parts, respectively, of the effective complex shear modulus $G^*_\mathrm{eff}(f)$ corresponding to the conditions shown in panels (a)-(c).}
    \label{fig:supp_panel_6_new}
\end{figure}

\subsection{Dependence of KLD on the order $m$ of autoregression model}
\label{sec:KLD}
\begin{figure}[H]
    \centering
    \includegraphics[scale=0.6]{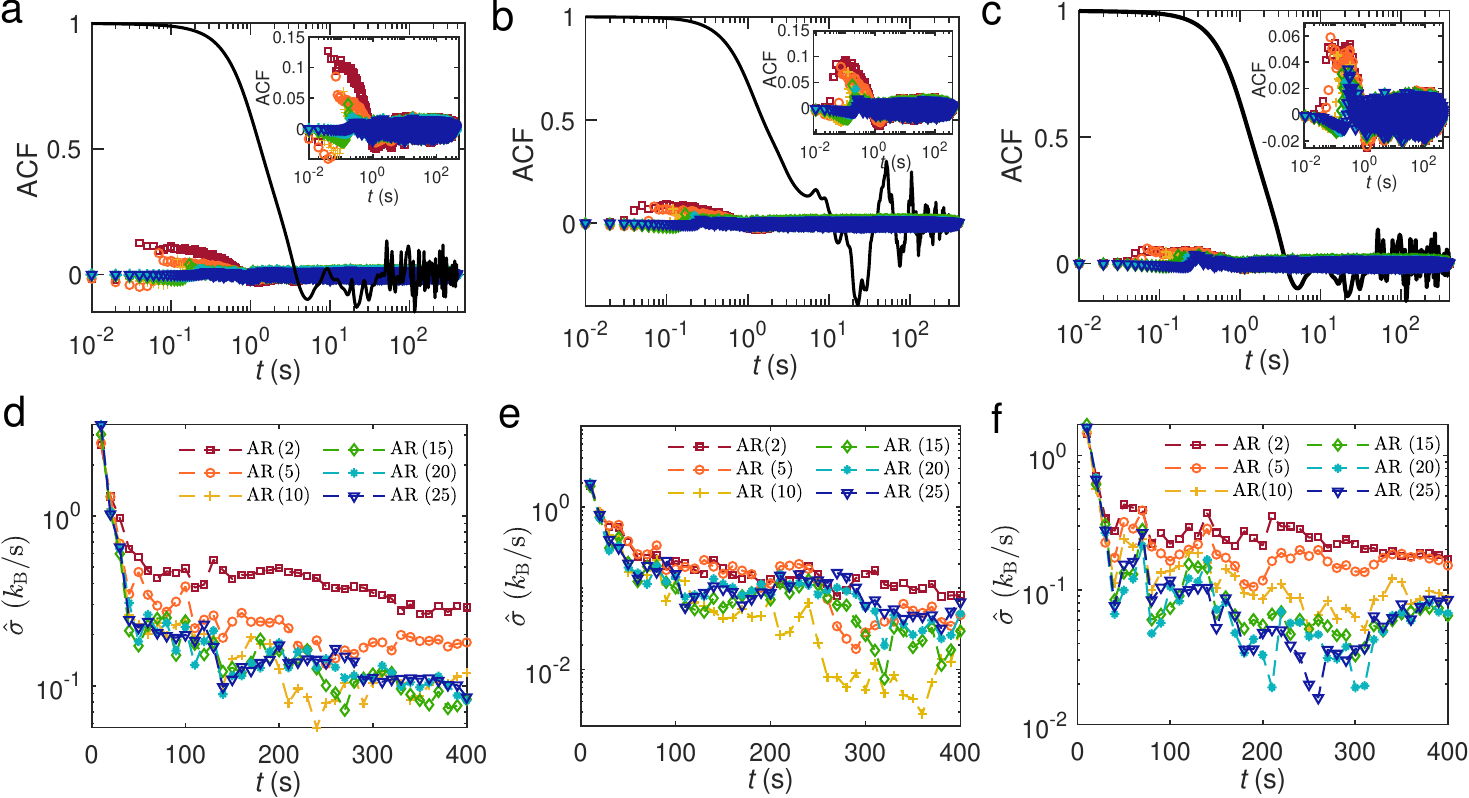}
    \caption{(a)–(c) Representative autocorrelation functions for three trajectories of the embedded AFM tip within the cortex of mitotically arrested HeLa cells, shown as a function of time. The solid black curves in (a)–(c) represent the autocorrelation function of the measured forward time series, while the symbols denote the autocorrelation function of the corresponding residual time series. These residuals are obtained by subtracting the auto-regression model AR\((m)\) for various values of \(m\) (labelled as legends within (d)-(f)) from the measured forward time series. (d)–(f) Local entropy production rates in the cortical layer of the HeLa cells, calculated using KLD for various orders of the autoregression model. }
    \label{fig:supp_panel_7_new}
\end{figure}

\subsection{Model Equations used in simulations}
\label{sec:Sims}
Following the approach described in \cite{villamaina_fluctuation-dissipation_2009} for dynamics with memory, we model cantilever motion in the presence of active forces in a viscoelastic environment by integrating the following system of 1st order differential Langevin equations 
\begin{eqnarray}
    \dot \delta_\mathrm{c}(t) &=& -\frac{(\kappa + \gamma_1/\tau)}{\gamma_\infty} \delta_\mathrm{c}(t)- \frac{\gamma_1}{\gamma_\infty\tau}\Delta(t) +\frac{\xi_1(t)}{\gamma_\infty} + \frac{F_{\rm cd}(t)}{\gamma_\infty}+\frac{F_{\rm burst}(t)}{\gamma_\infty}  \quad, \label{eq:numerical_model1}\\
     \dot \Delta(t) &=& -\frac{\delta_\mathrm{c}(t)}{\tau} -\frac{\Delta(t)}{\tau} +\frac{\xi_2(t)}{\gamma_1}  \quad, \label{eq:numerical_model2}\\
    \dot F_{\rm cd} (t)&=&  - \frac{F_{\rm cd} (t)}{\tau_{\rm cor}} + \sqrt{2 D_{\rm cage}} w(t) \quad, \label{eq:numerical_model3}\\
    F_{\rm burst}(t)&=& \sum_{i=1}^{N_{u}} \mathcal{A}_u \frac{e^{-(t-t_i)/\tau}}{\tau}\Theta(t-t_i) -\sum_{j=1}^{N_{d}} \mathcal{A}_d \frac{e^{-(t-t_j)^2/(2\sigma^2)}}{\sqrt{2\pi}\sigma}  \quad. 
    \label{eq:numerical_model4}
\end{eqnarray}
As described below, all parameters of the model are determined by experiments.
In particular, $\xi_1(t)$ and $\xi_2(t)$ are stochastic forces due to upscaled thermal noise according to measured frequency-dependent effective temperatures. Therefore, following \cite{jawerth_protein_2020}, we demand
$$
<\hat\xi_1(\omega)\hat\xi_1(\omega')>=4\pi k_\mathrm{B} T_{\rm eff}(\omega)\gamma_\infty\delta(\omega+\omega') \quad,<\hat\xi_2(\omega)\hat\xi_2(\omega')>=4\pi k_\mathrm{B} T_{\rm eff}(\omega)\gamma_1\delta(\omega+\omega') \quad, 
$$
where $T_{\rm eff}(f)=\left((f_{\rm act}-1)\exp(-2\pi f \tau_{\rm eff})+1\right)T$ is a frequency-dependent increased temperature. To approximate measured values of $T_{\rm eff}(f)$, see Fig.~\ref{fig:panel_2}c main text, we chose $f_{\rm act}=90, 90$ and $15$ to model the case of control, Latrunculin-treated and Blebbistatin-treated cells, respectively. For all conditions, we used the same timescale $\tau_{\rm eff}=2/3$~s.

$\Delta(t)$ is an auxiliary time variable that allows to take into account the viscoelastic nature of the mechanical environment consisting of the cortex modeled as a Jeffrey's fluid and the cantilever spring modeled as a Kelvin-Voigt element, see Fig.~\ref{fig:panel_4}(a) main text.  The timescale $\tau$ is the measured relaxation time of the viscoelastic Jeffrey's model, see Fig.~\ref{fig:panel_4}\,(a) main text and Fig.~\ref{fig:supp_panel_5_new}.  

Additionally, time-dependent diffusion of the cell surface, driven e.g. by fluctuations in cortical tension, is modeled as an active force $F_{\rm cd}$ simulated via the diffusion of a caging potential. Correspondingly, $w(t)$ is a normalized white Gaussian noise. The associated relaxation timescale, $\tau_{\rm cor} = 50$~s, is set based on the turnover of myosin proteins in the cortex \cite{ruffine2023twofold}. We chose $D_{\rm cage}=f_{\rm act}\times 10^{-3}$~N/(m$\cdot$s) $\times k_\mathrm{B} T$ to match the slope of the measured MSD at large timescales. 

Further, $F_{\rm burst}(t)$ is an active force capturing burst noise with positive and negative bursts mimicking corresponding force spikes and dips in the experimental cantilever height trajectories where positive bursts are clearly asymmetric, while negative bursts are more symmetric, see Fig.~\ref{fig:supp_panel_1_new}. We interpret these experimentally observed force spikes and dips as events of concerted activation of myosin mini-filaments  (force spikes) and sudden cortical material failure (force dips), respectively. Correspondingly, we  model a positive burst by a sudden increase  up to amplitude $\mathcal{A}_u$ followed by an exponential relaxation with decay time $\tau$. 
Further, a negative burst is modeled as normalized Gaussian with amplitude $\mathcal{A}_d$ and standard deviation $\sigma$. Positive and negative bursts are placed at random burst times $\{ t_i\}_{i=1}^{N_u}$ and $\{ t_j\}_{j=1}^{N_d}$, respectively,  such that on average a burst frequency of $f_{\rm u}$ and $f_{\rm d}$ is achieved for either of them. 
Motivated by experimental characterization, see Fig.~\ref{fig:supp_panel_1_new}, parameter values were chosen for control, Latrunculin-treated and Blebbistatin treated cells as $\mathcal{A}_u \in \{10, 5, 3\} \times 65$~pN$\times$s, $\tau=1.5~$s, $\mathcal{A}_d\in \{5, 4, 2.5\}\times 40$~pN$\times$s, $\sigma=0.5$~s, $f_u=8/400~$Hz, $f_d\in\{4, 4, 7\}/400$~Hz. 

Further, according to our rheological characterization, see Fig.~\ref{fig:supp_panel_6_new}, we chose $\eta_0\in \{3500, 2500, 500\}~$Pa$\cdot$s, 
$\eta_\infty\in \{35, 15,  0\}~$Pa$\cdot$s. 
In addition, $\gamma_\infty=\eta_\infty f_c +\gamma_{\rm bare}$ where $\gamma_{\rm bare}=0.005~N\cdot$s/m is the approximate friction coefficient associated to the fluctuations of the free cantilever, see Fig.~\ref{fig:panel_1} main text, and $f_c=1~\mu$m is the approximate proportionality coefficient $(6\tan(\beta)\delta_0/\sqrt{2})$  of AFM cantilever indentation of a pyramidal tip with opening angle to edge $\beta=17.5^\circ$ and indentation depth $\delta_0\approx 1~\mu$m. Analogously, the second dashpot in the Maxwell element of the Jeffrey's fluid is  $\gamma_1=(\eta_0-\eta_\infty)\times f_c$.

Numerical integration was achieved through a simple Euler forward method with time step $\Delta t=10^{-4}$~s. 
\begin{figure}[H]
    \centering
    \includegraphics[scale=0.65]{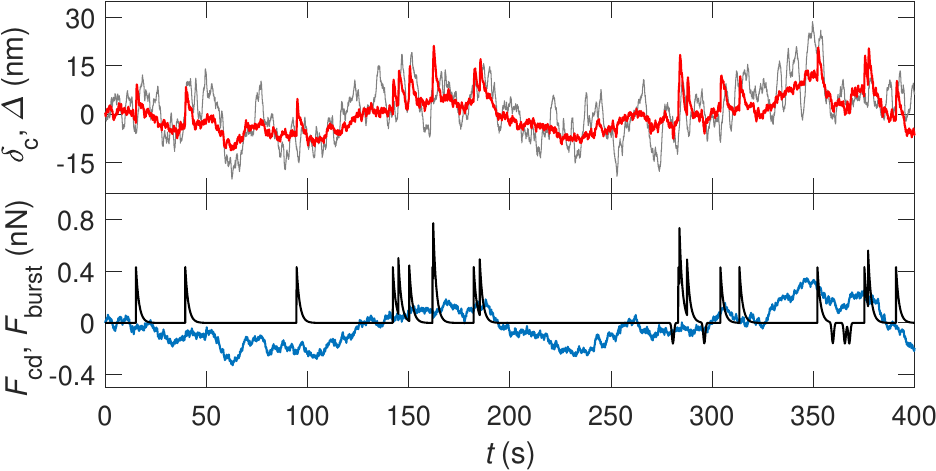}
    \caption{An exemplary numerical trajectory of the tip coordinate $\delta_\mathrm{c}$ (red curve) and the auxiliary coordinate $\Delta$ (gray curve), simulated using the model described by Eqns.~\eqref{eq:numerical_model1}-\eqref{eq:numerical_model4} (Upper panel). The lower panel shows the corresponding time evolution of the active Ornstein–Uhlenbeck force $F_\mathrm{a}$ (blue curve) and the burst force $F_\mathrm{burst}$ (black curve).}
    \label{supp_panel_8_new}
\end{figure}

\begin{figure}[H]
    \centering
    \includegraphics[scale=0.6]{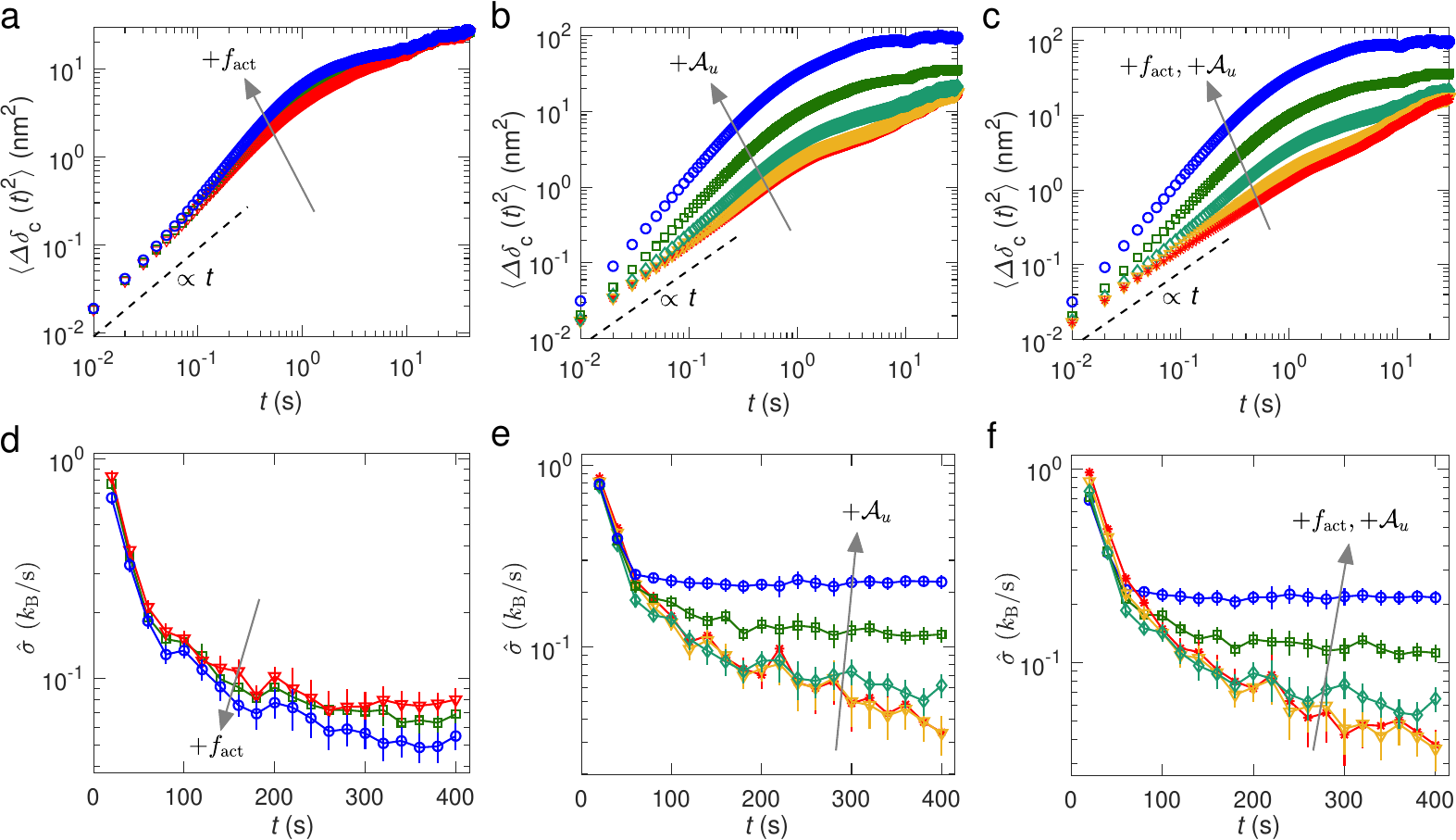}
\caption{
Influence of thermal noise strength and force spike amplitudes on MSD and corresponding KLD of simulated position time series. 
(a) MSD of the AFM tip, simulated using the model described by Eqns.~\eqref{eq:numerical_model1}-\eqref{eq:numerical_model4}, for different degrees of upscaling of the thermal noise amplitudes using increasing scaling factors $f_{\rm act}=20, 80$ and $320$. We conclude that upscaling thermal noise enhances the MSD.
(b) Numerically simulated mean squared displacements of the AFM tip for  increasing force spike amplitudes $\mathcal{A}_u\in 130, 260, 520, 1040$ and $2080$~pN$\times$s in the active force $F_\mathrm{burst}$.  We conclude that upscaling burst amplitudes enhances the MSD at large time scales.
(c) Numerically simulated mean squared displacements of the AFM tip assuming joint upscaling of thermal noise amplitudes $f_{\rm act}$ and force spike amplitudes $\mathcal{A}_u$ (parameters: $f_{\rm act}=20, 40, 80, 160$ and $320$ and $\mathcal{A}_u=130, 260, 520, 1040$ and $2080$~pN$\times$s). 
(d) Time irreversibility measure  $\hat{\sigma}$ over observation time for the trajectories corresponding to the MSDs shown in (a), with matching symbols and color codes. We conclude that upscaling thermal noise reduces the KLD.
(e) Time irreversibility measure $\hat{\sigma}$ over observation time for the trajectories associated with the MSDs in (b), using the same symbols and color codes.  We conclude that upscaling burst amplitudes enhances also the KLD.
(f) Time irreversibility measure $\hat{\sigma}$ over observation time for the trajectories associated with the MSDs in (c). 
We conclude that upscaling thermal noise and time asymmetric bursts jointly leads to a net increase of $\hat\sigma$.
In all cases ((a)–(f)), MSDs and $\hat{\sigma}$ are computed by averaging over 15 independent trajectories. If not indicated otherwise, parameters were chosen using the 'control' parameter set. 
}
\label{fig:supp_panel_9_new}

\end{figure}
\begin{figure}[H]
    \centering
    \includegraphics[scale=0.62]{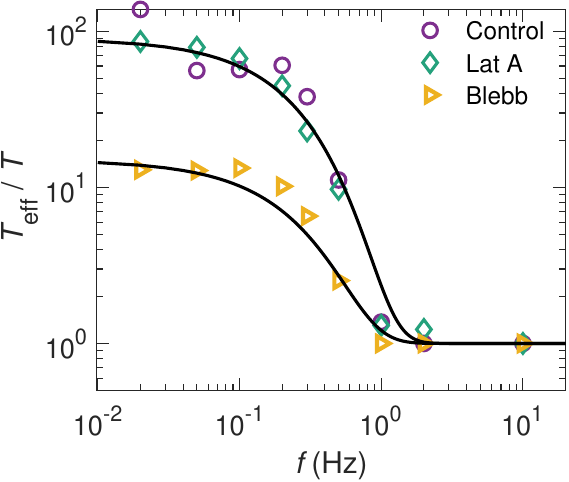}
    \caption{Effective temperature $T_{\text{eff}}$ normalized by the ambient temperature $T$ as a function of frequency, determined from the violation of the fluctuation-dissipation theorem (FDT). Data points represent control cells (purple circles), cells treated with Latrunculin-A (green diamonds), and cells treated with Blebbistatin (yellow triangles). The solid curves show the corresponding analytical approximation to the function $T_{\text{eff}}/T = (f_{\rm act} - 1)\exp(-2\pi f \tau_{\text{eff}})+1$ used in the minimal model  with $\tau_{\text{eff}} = 2/3~\mathrm{s}$ for all cases. The parameter $f_{\rm act}$ is 90 for control and Latrunculin-treated cells, and 15 for Blebbistatin-treated cells.}
    \label{fig:supp_panel_10_new}
\end{figure}
\end{suppenv}

\end{document}